\documentclass[column,secnumarabic,amssymb,amsmath,nofootinbib,floatfix,
nobibnotes,aps,
preprint,
superscriptaddress,
prd
]{revtex4}

\usepackage{caption}
\usepackage{amsmath,adjustbox,mathtools,amssymb}
\usepackage{amsfonts}
\usepackage{amssymb}
\usepackage{graphicx}
\usepackage{braket}
\usepackage{slashed}
\usepackage{wrapfig}
\usepackage{tikz-feynman,contour}
\usepackage{tikz-feynhand}
\usepackage{feynmp-auto}
\usepackage{bbold}
\usepackage{bm}
\usepackage{verbatim}
\usepackage{subcaption}
\usepackage{appendix}
\usepackage{slashed}
\usepackage{ulem}
\normalsize
\usepackage{booktabs}
\usepackage{supertabular}
\usepackage{tabularx}
\usepackage{tabulary}
\usepackage{multirow}
\usepackage{soul}

\newcommand{\ud}{{\rm{d}}}

\newenvironment{system}%
{\left\lbrace\begin{array}{@{}l@{}}}%
{\end{array}\right.}

\begin{document}

\title{Two-particle two-hole excitations in semi-inclusive  neutrino-nucleus scattering}
\author{V. Belocchi} 
\affiliation{
	Dipartimento di Fisica, Universit\`a di Torino, via P. Giuria 1, 10125 Turin, Italy}
\affiliation{
	INFN, Sezione di Torino, via P. Giuria 1,10125, Turin, Italy
}
\affiliation{
Instituto de Física Corpuscular (IFIC), Consejo Superior de Investigaciones Científicas (CSIC) and Universidad de Valencia, E-46980 Paterna, Valencia, Spain
}\author{M.B. Barbaro} 
\affiliation{
	Dipartimento di Fisica, Universit\`a di Torino, via P. Giuria 1, 10125 Turin, Italy}
\affiliation{
	INFN, Sezione di Torino, via P. Giuria 1,10125, Turin, Italy
}
\author{A. De Pace} 
\affiliation{
	INFN, Sezione di Torino, via P. Giuria 1,10125, Turin, Italy
}
\author{M. Martini}
\affiliation{IPSA-DRII,  63 boulevard de Brandebourg, 94200 Ivry-sur-Seine, France}
\affiliation{Sorbonne Universit\'e, CNRS/IN2P3, Laboratoire
de Physique Nucl\'eaire et de Hautes Energies (LPNHE), Paris, France}
\begin{abstract}
A calculation of the two-particle two-hole contribution to the semi-inclusive  $(\nu_\mu,\mu^-N)$ cross section on carbon is performed in the framework of the relativistic Fermi gas model. The process is  driven by meson-exchange currents encompassing all contributions involving the exchange of a single pion and the excitation of a $\Delta$ resonance. The calculation is validated against the inclusive $(\nu_\mu,\mu^-)$ response functions already existing in the literature, and then extended for the first time to the semi-inclusive channel. Results are presented both at fixed neutrino energy and folded with the T2K neutrino flux. 
\end{abstract}

\maketitle
\section{Introduction}

Understanding and accurately describing 
the properties of neutrinos is one of the primary and most challenging goals of the neutrino physics community,
nowadays especially focused in the precise determination of the PMNS oscillation parameters. Particular attention is paid to the CP violating Dirac phase, the main purpose of the current long-baseline accelerator experiments such as T2K and NO\(\nu\)A, as well as future projects like DUNE and HK. With sufficient accuracy, such experiments  also aim to determine the mass hierarchy of the three neutrino massive eigenstates. 

The portal to the precision era is a deeper comprehension of the nuclear interaction with an external probe and, more generally, of the nuclear many-body system. Due to the very low interaction strength of neutrinos with matter, achieving the necessary statistics requires the use of nuclear targets.  As a results, accurate nuclear physics input is essential for the success of current and future neutrino experiments.

In long-baseline experiments, the knowledge of the initial neutrino energy $E_\nu$ is crucial for the extraction of the oscillation parameters. However, the incident flux is widely distributed in energy, and $E_\nu$ must be reconstructed from the detected final state.
This demands a consistent theoretical description of all active interaction channels at the energy scales of the incoming beam. Among these, the two-particle two-hole ($2p2h$) channel, corresponding to the emission of two nucleons from the target nucleus, is one of the most challenging to evaluate, since it involves the coupling of the probe to a two-body current.

The realization of the importance of the $2p2h$ channel for neutrino experiments dates back to 2009-2010, when the first measurement of the double differential QE-like cross section (no pions in the final state) performed by the MiniBooNE experiment~\cite{MiniBooNE:2010bsu} revealed a strong disagreement with the prediction based on the Fermi gas model, implemented in all major Monte Carlo event generators.
The explanation of this disagreement was first suggested by the authors of Refs.~\cite{Martini:2009uj,Martini:2010ex}, who showed that
 the inclusion of \(2p2h\) could explain the MiniBooNE data. It is now well established that \(2p2h\) excitations provide a significant contribution to the inclusive $(\nu_l, l)$ cross section without pions in the final state. Several calculations of this contribution are available~\cite{Martini:2009uj,Nieves:2011pp,RuizSimo:2016rtu,VanCuyck:2017wfn,Rocco:2018mwt,Lovato:2020kba}, all generally consistent with the inclusive experimental data within the error bars. However, the spread between different theoretical predictions is still quite large and should be reduced in order to meet the desired precision.

Further constraints on the nuclear modeling of neutrino cross sections can be obtained through comparison with semi-inclusive measurements, where one or more hadrons in the final state are detected in coincidence with the final charged lepton. These observables are significantly more sensitive to the details of the nuclear model than inclusive observables. Several such measurements have recently been carried out \cite{T2K:2018rnz,MINERvA:2018hba,MINERvA:2019ope,MicroBooNE:2020fxd,MicroBooNE:2020akw,MicroBooNE:2023tzj}. 
This has led to a great demand for theoretical models capable of providing predictions for the cross section as a function of the hadronic variables.  A considerable portion of the theoretical physics community has focused its efforts in this direction, given that the vast majority of available interaction models were designed to reproduce data from strictly inclusive measurements, integrating over the entire hadronic part of the process. 
For example, recent efforts have investigated the semi-inclusive reaction $(\nu_l,lN)$ in the QE channel, corresponding to the scattering of the probe with a single nucleon~\cite{Moreno:2014kia,Franco-Patino:2022tvv,Franco-Patino:2023msk}. 

 The situation is markedly different for semi-inclusive \(2p2h\) processes. A fully microscopic calculation of the contribution of semi-inclusive reaction $(\nu_l, lN)$ in the \(2p2h\) channel is still missing.
  So far the only available result in this direction comes from the Ghent group \cite{VanCuyck:2016fab,VanCuyck:2017wfn}, which, however, does not include the $\Delta$ contributions to meson-exchange currents. Semi-inclusive $(\nu_l,lNN)$ cross sections at fixed neutrino energies have also recently been computed in  \cite{Martinez-Consentino:2023hcx,sym16020247}. 
  
   Currently the \(2p2h\) component of the cross section is simulated in event generators on the basis of {\it inclusive} calculations \cite{Nieves:2011pp,RuizSimo:2016rtu}, following a procedure that necessarily relies on some strong approximations. Inclusive processes involve a summation over all accessible intermediate nuclear states, consequently predicting cross sections only as functions of the outgoing lepton kinematics. In principle, such approach cannot be employed to predict semi-inclusive or exclusive observables, where a specific hadronic final state is detected. 
    However, 
given the lack of theoretical calculations of these contributions, the strategy taken so far has been to ``extract'' exclusive predictions from inclusive results, forcibly using assumptions \cite{Dolan:2019bxf, Hayato:2021heg} whose reliability is difficult to control. 

Nonetheless, \(2p2h\) contributions represent a significant component of the semi-inclusive cross section, as shown in Refs.~\cite{Franco-Patino:2022tvv, Franco-Patino:2023msk} for neutrinos and in Ref.~\cite{CLAS:2021neh} for electrons. Consequently, not only is the implementation of the \(2p2h\) itself questionable, but it also has implications for the conclusions drawn concerning nuclear models used to describe the one-body nuclear response.
For instance, the $(\nu_\mu, \mu p)$ cross section, recently measured by T2K \cite{T2K:2018rnz}, MINERvA \cite{MINERvA:2019ope}, and MicroBooNE \cite{MicroBooNE:2020fxd}, has been shown to be highly sensitive to the model used to describe the final state interactions (FSI) between the ejected proton and the residual nucleus. Different FSI prescriptions can yield significantly different results in general \cite{Franco-Patino:2022tvv, Franco-Patino:2023msk}, and the comparison of theoretical predictions with data is strongly influenced by the contribution of \(2p2h\) events to the experimental signal.
The only correct approach to implement a model for \(2p2h\) in a Monte Carlo generator for the semi-inclusive reaction is through a fully microscopic calculation. 
The main scope of this work is to fill this gap,  providing the neutrino physics community with a model tailored for semi-inclusive 
\(2p2h\) contributions. The calculation presented here extends the electromagnetic $(e,e'p)$ formalism developed in Ref.~\cite{Belocchi:2024rfp} to the weak interaction sector.

The paper is organized as follows: In Sect. \ref{sec:Formalism} we introduce the formalism for semi-inclusive neutrino-nucleus scattering, with particular emphasis on the $2p2h$ channel; in Sect. \ref{sec:Model} we describe the nuclear model adopted in this work; in Sect. \ref{sec:Results} we report  numerical results for both inclusive and semi-inclusive neutrino-nucleus cross sections, and in Sect. \ref{sec:Concl} we draw our conclusions and present future developments.

\section{Formalism}
\label{sec:Formalism}

In this Section, the formalism developed in Ref.~\cite{Belocchi:2024rfp} for semi-inclusive electron-nucleus scattering is extended to describe the charged-current (CC) semi-inclusive neutrino-nucleus interactions.

In a semi-inclusive process one or more hadrons in the final state are detected in coincidence with the outgoing lepton. 
\begin{figure}[ht!]
    \centering
\includegraphics[width=.4\linewidth]{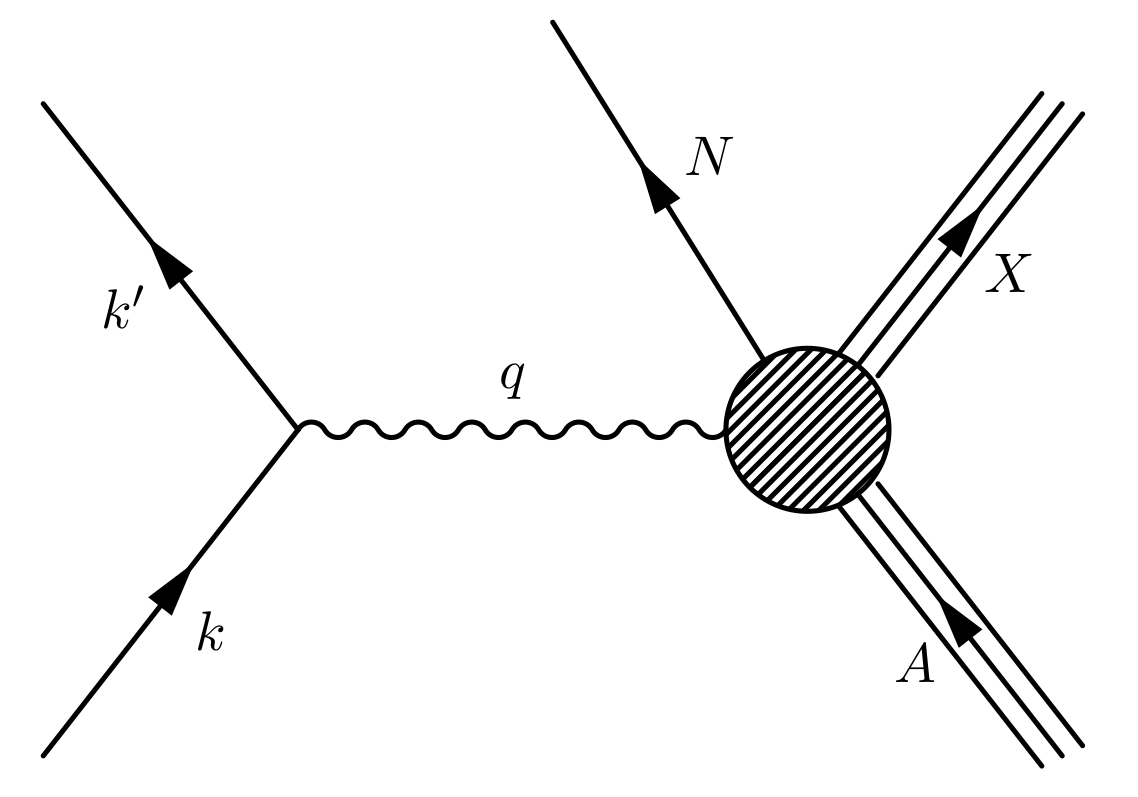}
\caption{Lepton-nucleus scattering in which a final nucleon \(N\) is detected in coincidence with the outgoing lepton. \(X\) remains the unobserved hadronic final state, not including \(N\).}
    \label{fig:l-A-semi-scatter}
\end{figure}
We focus here on the specific case depicted in Fig.~\ref{fig:l-A-semi-scatter}, in which a muon neutrino of momentum $k$ scatters off a nucleus $A$ at rest in the laboratory frame via the charged current process:
\begin{equation}
\nu_\mu+A \to \mu^-+N+X \,.
\label{eq:eepN}
\end{equation}
  The final state comprises the muon \(\mu^-\),  with momentum $k'$ and solid angle $\Omega_{k'}$, a knocked-out nucleon, with momentum $p_N$ and solid angle $\Omega_{N}$, and the residual system $X$. 
Without loss of generality, we define the scattering plane, spanned by the incoming and outgoing lepton momenta, as the \(xz\) plane. 
In contrast to inclusive processes, where the hadronic final states are unobserved and an integration over all such configurations is performed, semi-inclusive observables retain sensitivity to specific hadrons in the final state. 

The six-fold differential semi-inclusive cross section, with respect to the muon and nucleon variables, takes the general form:
\begin{equation}
    \frac{\rm d^6 \sigma}{\rm d E_{k'}\rm d \Omega_{k'} \ud E_{N} \ud \Omega_{N}}=\sigma_{0} p_N E_N \mathcal{F}_{N}^2\,,
    \label{eq:xsec-semi1}
\end{equation}
where \begin{equation}
    \sigma_0=\frac{G_F^2}{8\pi^2}\cos^2\theta_C\frac{|\bf{k'}|}{E_k}\nu_0\, ,
    \label{eq:sigma0EW}
\end{equation}
$G_F=g_W^2/4\sqrt{2}M_W^2$ being the Fermi constant, $\theta_C$ the Cabibbo mixing angle and 
    \begin{equation}
    \nu_0\equiv(E_k+E_{k'})^2-|{\bf q}|^2=4E_k E_{k'} -Q^2 \,,
\end{equation} 
with \(q^\mu=(\omega,{\bf q})=k^\mu-k'^\mu\) the momentum transfer, {\it{id est}} the \(W\)-boson momentum, and  $Q^2\equiv-q^2>0$. 
The quantity 
\begin{equation}
    \mathcal{F}_N^2=\frac{\widetilde L_{\mu \nu}}{\nu_0}W^{\mu \nu}_{A(N)}\,,
    \label{F2def}
    \end{equation}
involves a contraction between the weak lepton tensor
\begin{equation}
\widetilde L_{\mu\nu}=k_\mu k'_\nu+k'_\mu k_\nu-k\cdot k' g_{\mu\nu}+i \epsilon_{\mu\nu\rho\sigma} k^\rho k'^\sigma
\end{equation}
and the semi-inclusive weak nuclear tensor \cite{Belocchi:2024rfp}
\begin{equation}
    W^{\mu\nu}_{A(N)} = \sum_X \langle A|\hat J^{\mu\dagger}|N,X\rangle \langle N,X| \hat J^\nu | A\rangle
    \,\delta\left(E_{N}+E_X-E_0-\omega\right) \,.
    \label{eq:WmunuAN}
\end{equation}
The latter 
encompasses all the nuclear properties, such as initial- and final-state interactions between nucleons, nuclear recoil, absorbed energy, reflecting all the features of the adopted nuclear model.
In Eq.~\eqref{eq:WmunuAN}, 
$|A\rangle$ denotes the nuclear ground state of energy $E_0$, the ket $|N,X\rangle$ is the hadronic final state of energy $E_N+E_X$ and a sum over the unobserved states $X$ is performed. 
The weak nuclear current $\hat J^\mu =  \hat J^\mu_{1b} +\hat J^\mu_{2b}+...+\hat J^\mu_{Ab} $ is the sum of one-, two-, and many-body currents, with the last term $\hat J^\mu_{Ab}$ involving all the  \(A\) nucleons.
As a consequence, the hadronic tensor can be expressed as the sum of various contributions corresponding to the excitation of different final states. Here we shall retain only the one- and two-body currents, which can excite 
one-particle-one-hole (\(X=1p1h\)) and 
two-particle-two-hole (\(X=2p2h\)) states:
\begin{equation}
    W^{\mu\nu}_{A(N)} \simeq W^{\mu\nu}_{A(N), 1p1h}+W^{\mu\nu}_{A(N),2p2h}\,,
    \label{eq:W1+W2}
\end{equation}
and we shall focus in particular on $W^{\mu\nu}_{A(N),2p2h}$.
Note, however, considering a description of the lepton-nucleus scattering without taking into account the medium corrections, -for instance, computing the self energy of the particles in the nucleus- the two-body current contributes also to the 1p1h tensor through interference with the one-body current~\cite{Casale:2025wsg,Casale:2025avv}.

With respect to the electromagnetic case studied in Ref.~\cite{Belocchi:2024rfp},
 the axial current introduces additional  contributions. The non-conservation of the axial current and the vector-axial interference add complexity to neutrino CC semi-inclusive scattering, making it significantly more intricate than electron scattering.  Moreover, compared to the inclusive case and coherently with the electromagnetic case, new semi-inclusive contributions arise because the azimuthal angle invariance no longer holds. This results in the presence of ten distinct non-vanishing response functions:
\begin{eqnarray}
\mathcal{F}^2_{N}&=&V_{CC}R_{CC}^{(N)}-2V_{CL}R_{CL}^{(N)}+V_{LL}R_{LL}^{(N)}+V_{T}R_{T}^{(N)} + 2V_{T'}R_{T'}^{(N)}
\nonumber\\
 &-& V_{CT} R_{CT}^{(N)}+V_{LT} R_{LT}^{(N)} +V_{TT} R_{TT}^{(N)} - V_{C\bar T'} R_{C\bar T'}^{(N)} - V_{L\bar T'} R_{L\bar T'}^{(N)}\,.
 \label{eq:F2}
\end{eqnarray}
Here, the kinematic factors $V_K$ and the response functions $R_K^{(N)}$ are combinations of leptonic and hadronic tensor components, respectively. The explicit expressions are given in Appendix \ref{app:resp}.

In the next Section we define the model adopted to evaluate the $2p2h$ nuclear tensor, specifying  the nuclear model and the two-body currents.

\section{Model}
\label{sec:Model}

To compute the $2p2h$ hadronic tensor, we adopt as wave-function basis the Relativistic Fermi Gas (RFG) model, which treats the nucleus as a system of non-interacting nucleons described by Dirac spinors, correlated only by the Pauli exclusion principle.
 
It is worth mentioning that, in the QE channel, the RFG model is inadequate to provide semi-inclusive predictions. This is a consequence 
of the too simple description of the lepton-nucleus interaction process, associated in this case to a one-body current. 
The nuclear semi-inclusive tensor in the RFG framework entails a three-dimensional integration over the possible initial nucleon states. The four-momentum conservation simply reduces the integration, implying that, for a given nucleon observed in the final state, only one initial nucleon configuration is allowed. This is far from being realistic, but is expected from the 
momentum distribution provided by the RFG, which is a step function: $n(p)=\theta(p_F-p)$. This failure is mostly due to the infinite nuclear matter approximation underlying the model.

On the other hand, in the $2p2h$ channel considered in this work the lepton-nucleus interaction is carried by a two-body current acting on two different nucleons in the Fermi sphere, which introduces dynamical correlations between the nucleons. So, while the calculation is performed on the basis of the RFG model, the nuclear tensor goes beyond the pure Fermi gas approximation.

The RFG-based \(2p2h\) nuclear tensor is given by:
\begin{equation}
\begin{aligned}
  W_{A(N)\,2p2h}^{\mu \nu}=& (2\pi)^3V  \frac{m_N}{(2\pi)^3E_{p_1}}
  \theta(|\mathbf{p_1}|-p_F|)
\left[ \prod_{i=1}^2 \int_{|\mathbf{h}_i|\leq p_F} \frac{m_N \, \ud \mathbf{h_i}}{(2\pi)^3E_{h_i} }\right] \int \frac{m_N \, \ud \mathbf{p_2}}{(2\pi)^3E_{p_2}} \theta(|\mathbf{p_2}|-p_F|)   \\
  &\times w_{2p2h}^{\mu \nu}(h_1,h_2,p_1,p_2)\, \delta^4(q +h_1+h_2 -p_1-p_2)\,,
  \end{aligned} 
  \label{eq:WmunuN}
  \end{equation} 
  where \(m_N\) is the nucleon mass and \(V\) the volume of the quantized nuclear system
  \begin{equation}
      V=(2\pi)^3\frac{3\mathcal{N}}{8\pi p_F^3}\,.
  \end{equation}
  In Eq. \eqref{eq:WmunuN} the nucleon detected has been arbitrarily chosen to be the one having momentum \(\mathbf{p_1}\), without loss of generality. 
  Pauli blocking is encoded in the two step functions acting on the final-particle momenta $\mathbf{p_1}$ and $\mathbf{p_2}$.
  It should be noted that in the semi-inclusive two-body tensor definition two final indistinguishable nucleons appear and can be associated to the momentum \(\mathbf{p_1}\). This fact is taken into account in the isospin computation part: if the detected particle is a proton, then every isospin configuration that admits at least one proton in the final state contributes to the semi-inclusive cross section, including those in which the proton state is associated with \(\mathbf{p_2}\). This is allowed thanks to the elementary tensor symmetry under the interchange of the final nucleons kinematical quantities.
  
The semi-inclusive \(2p2h\) elementary tensor appearing in Eq.~\eqref{eq:WmunuN} is defined as follows:
\begin{equation}
     w^{\mu\nu}_{2p2h}(h_1,h_2,p_1,p_2) = \frac{1}{4}      \sum_{\substack{\mathrm{spin}\\ 
    \mathrm{isospin*}}}\langle h_1h_2 |\hat J^{\mu\dagger}_{2b} | p_1p_2\rangle\langle p_1 p_2| \hat J^\nu_{2b} |h_1h_2\rangle  ,
    \label{eq:reduced_w}  
\end{equation}
where $\hat J^\mu_{2b}$ is the two-body current operator, while the kets \( \ket{h_1 h_2} \) and \( \ket{p_1 p_2} \) indicate the initial and final two nucleon states, respectively, characterized by their momenta.
The factor $1/4$ accounts for the $1/2$, for each bracket, that naturally emerges in the definition of a two-body operator acting on the hole pair, when a sum over all the possible initial states is performed: this procedure avoids double counting.
The notation $\sum_{\substack{\mathrm{spin}\\ 
    \mathrm{isospin*}}}$ indicates that the sum is performed on all possible isospin configurations that describe at least one particle of the detected isospin (proton or neutron), and the sum over the final isospin of the detected particle is not performed.
Due to the antisymmetry of the two-nucleon wave function, the matrix element present in Eq.~\eqref{eq:reduced_w} can be written as the difference of two terms:
\begin{align}
     \langle p_1 p_2 | \hat J^\mu_{2b} | h_1 h_2 \rangle
    &=j^\mu(h_1,h_2,p_1,p_2)-j^\mu(h_1,h_2,p_2,p_1) \,,
\end{align}
here called "{normal ordered}" (NO) and "{inverted ordered}" (IO).
The product of the two matrix elements appearing in Eq.~\eqref{eq:reduced_w} leads then to four terms. Thanks to 
symmetry properties, it is possible to group these terms in two different contributions, called "{direct}" and "{exchange}":
\begin{eqnarray}
    w^{\mu\nu}_{2p2h} = \frac{1}{4}\big[\underbrace{j^{\mu \dagger}_{\rm NO}\,j^{\nu}_{\rm NO}}_{\rm direct}+\underbrace{j^{\mu \dagger}_{\rm IO}\,j^{\nu}_{\rm IO}}_{\rm direct} -  \underbrace{j^{\mu \dagger}_{\rm NO}\,j^{\nu}_{\rm IO}}_{\rm exchange} - \underbrace{j^{\mu \dagger}_{\rm IO}\,j^{\nu}_{\rm NO}}_{\rm exchange}\big] \,.   
\end{eqnarray}
 In the calculation of the nuclear tensor \eqref{eq:WmunuN}, an integration over all possible initial states is performed. Due to the 
 structure of the two-body current, the tensor \(w^{\mu \nu}\) exhibits a symmetry under the exchange of the holes and particles variables, namely \(1 \leftrightarrow 2\). Then the two direct terms provide the same contribution and so do the exchange ones. This allows us to compute just two of the four terms that appear in the reduced tensor definition.

The two-body current operator we adopt to evaluate the tensor \eqref{eq:reduced_w} is a Meson Exchange Current (MEC),
obtained from a chiral Lagrangian which describes the interaction between nucleons, pions and the $\Delta$ resonance \cite{Scherer:2012xha}.
The current is the sum of several contributions:
\begin{equation}
\hat J^\mu_\mathrm{MEC}\equiv \hat J^\mu_\mathrm{pif}+\hat J^\mu_\mathrm{sea}+\hat J^\mu_\mathrm{pp}+\hat J^\mu_{\Delta_\mathrm{F}}+\hat J^\mu_{\Delta_\mathrm{B}}\,,
\end{equation}
where the first three operators, denoted as pion-in-flight (pif), seagull (sea) and  pion-pole (pp), do not involve the $\Delta$ resonance, which is present in two last contributions, the $\Delta$ forward and backward terms ($\Delta_\mathrm{F}$, $\Delta_\mathrm{B}$).

\begin{figure}[ht!]
    \centering
\includegraphics[width=.9\linewidth]{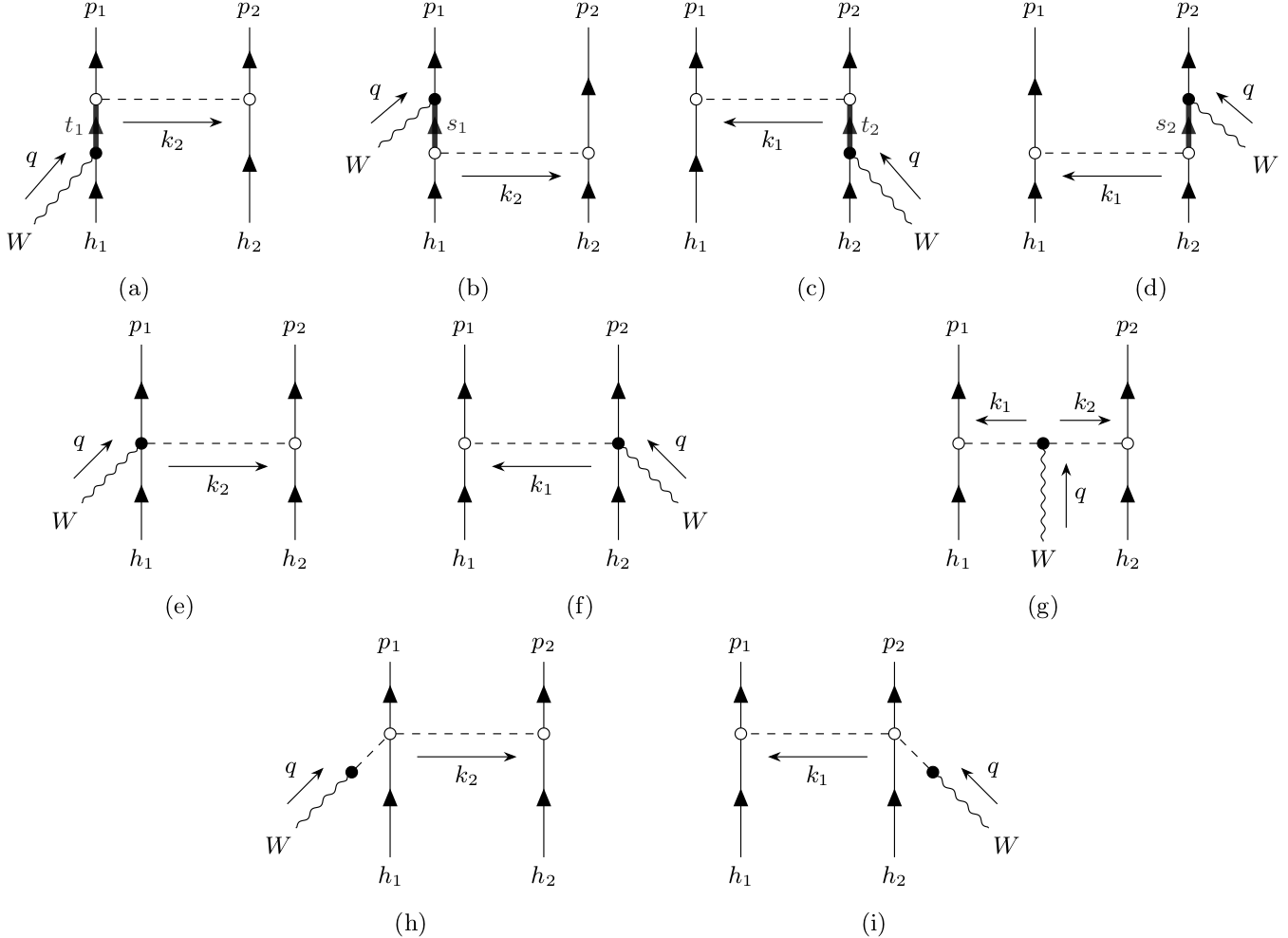}
\caption{First row: $\Delta$-MEC diagrams, forward (a,c) and backward (b,d). The $\Delta$ resonance is represented by the thick line. Second and third row: pure pionic diagrams, seagull (e,f), pion-in-flight (g) and pion-pole (h,i). The labels $h_i$ and $p_i$ stand for the initial and final nucleon states, respectively. The \(W\) vertex is the black solid dot, while internal nuclear vertices are the white ones. }
    \label{fig:2p2h} 
\end{figure}

By isolating the isospin operators that appear in each term, one can write the MEC matrix elements as
\footnote{The isospin operators are defined as: \begin{equation}
    \bm I_V\equiv i\big(\bm \tau^{(1)} \times \bm \tau^{(2)} \big)\qquad  I_{V_\pm}\equiv I_{V_1}\pm iI_{V_2} \, .
    \end{equation}
In terms of physical quantities, related to physical pions, the operators can be written as
\begin{equation}
    I_{V\pm}=\mp \big(\tau_3^{(1)}\tau_\pm^{(2)}-\tau_\pm^{(1)}\tau_3^{(2)} \big)\qquad \tau_\pm=\tau_1\pm\tau_2\;.
\end{equation}
The superscript \( ^{(i)}\) means that the operator acts on the  \(i\)-th particle of the pair.}
\begin{equation}
\langle p_1 p_2 | \hat J^\mu_\mathrm{MEC} | h_1 h_2 \rangle \equiv I_{V_\pm}\Big( j^\mu_\mathrm{pif}+j^\mu_\mathrm{sea}+j^\mu_\mathrm{pp}+j^\mu_{\Delta_3}\Big)+2\tau_\pm ^{(1)}j^\mu_{\Delta_1}+2\tau_\pm ^{(2)}j^\mu_{\Delta_2}\,, 
\end{equation}
where 
\begin{equation}
    \begin{aligned}       j_{\mathrm{pif}}^\mu&=\bigg(\frac{f_{\pi NN}}{m_\pi}\bigg)^24m_N^2 F_1^V(q^2) \Delta_\pi(k_1)\Delta_\pi(k_2)F_{\pi NN}(k_1)F_{\pi NN}(k_2) (k_1^\mu-k_2^\mu) \bar{u}_{p_1} \gamma_5 u_{h_1}\bar{u}_{p_2} \gamma_5 u_{h_2}\\
   j^\mu_{\mathrm{sea}_V} &=\bigg(\frac{f_{\pi NN}}{m_\pi}\bigg)^2 2m_N F_1^V(q^2) \Big[\Delta_\pi(k_1) F_{\pi NN}^2(k_1)
        \bar{u}_{p_1} \gamma_5 u_{h_1} \bar{u}_{p_2} \gamma^\mu \gamma_5 u_{h_2} - (1 \leftrightarrow 2) \Big]\\
  j^\mu_{\mathrm{sea}_A}&= \bigg(\frac{f_{\pi NN}}{m_\pi}\bigg)^2\frac{1}{g_A}2m_N \Big[ F_\rho(k_2^2)\Delta_\pi(k_1) F_{\pi NN}^2(k_1)\bar{u}_{p_1} \gamma_5 u_{h_1} \bar{u}_{p_2} \gamma^\mu \gamma_5 u_{h_2}  - (1 \leftrightarrow 2) \Big]\\
  j^\mu_{\mathrm{pp}} &= \bigg(\frac{f_{\pi NN}}{m_\pi}\bigg)^2\frac{1}{g_A}2m_N \Delta_\pi(q) q^\mu \Big[F_\rho(k_2^2)\Delta_\pi(k_1) F_{\pi NN}^2(k_1)\bar{u}_{p_1} \gamma_5 u_{h_1}   \bar{u}_{p_2} \slashed q u_{h_2} - (1 \leftrightarrow 2) \Big]\\
    j^\mu_{\Delta_1}&= -\frac{1}{\sqrt{6}}\frac{f^*}{m_\pi}\frac{f_{\pi NN}}{m_\pi}2m_N F_{\pi N\Delta}(k_1)\Delta_\pi(k_1)F_{\pi NN}(k_1)\bar{u}_{p_1} \gamma_5 u_{h_1} \,\\ 
        & \qquad \qquad \qquad \qquad  \qquad\qquad \quad \cdot\bar{u}_{p_2}k_1^\alpha \Big[G_{\alpha \beta}(t_2)\Gamma^{\beta \mu}(h_2,q)+ \tilde \Gamma^{ \mu \beta}(p_2,q)G_{\beta \alpha}(s_2)\Big]   u_{h_2}\\
    j^\mu_{\Delta_2}&= j^\mu_{\Delta_1}(1\leftrightarrow 2)  \phantom{\frac{f_{N}}{\sqrt{6}}}\\
    j^\mu_{\Delta_3}&= -\frac{1}{\sqrt{6}}\frac{f^*}{m_\pi}\frac{f_{\pi NN}}{m_\pi}2m_N\bigg\{ F_{\pi N\Delta}(k_1)\Delta_\pi(k_1)F_{\pi NN}(k_1)\bar{u}_{p_1} \gamma_5 u_{h_1} \,\\ 
        &   \qquad \qquad  \qquad \qquad\cdot \bar{u}_{p_2}k_1^\alpha \Big[G_{\alpha \beta}(t_2)\Gamma^{\beta \mu}(h_2,q)- \tilde \Gamma^{ \mu \beta}(p_2,q)G_{\beta \alpha}(s_2)\Big]  u_{h_2} -(1\leftrightarrow 2)  \bigg\}\, 
    \end{aligned}
    \label{eq:MEC}
\end{equation} 
correspond to the Feynman diagrams shown in Fig.~\ref{fig:2p2h}.

In the above expressions, $k_i=p_i-h_i$ is the four-momentum carried by the pion. Moreover, the form factors $F_1^V$, $F_{\pi NN}$ and $F_{\pi N\Delta}$, and the pion and $\Delta$ propagators,  $\Delta_\pi$ and $G_{\alpha\beta}$, have been introduced. Their expressions can be found in Appendix \ref{app:FF}. 
The form factor $F_{\rho}$ is associated to the $\rho$ exchange in the vector contact $W N N \pi$ interaction, following the vector-meson dominance assumption. The $WN\Delta$ transition operator
$\Gamma^{\alpha\beta}$ is defined as:
\begin{equation}    \Gamma^{\alpha\beta}=\Gamma_V^{\alpha\beta}+\Gamma^{\alpha\beta}_A
\end{equation}
with
\begin{eqnarray}
    \Gamma^{\alpha \mu}_V(p,q)&=&\bigg[\frac{C_{3V}}{m_N}(g^{\alpha\mu}\slashed q -q^\alpha \gamma^\mu)+\frac{C_{4V}}{m_N^2}(g^{\alpha \mu}q\cdot p_\Delta-q^\alpha p_\Delta^\mu)    +\frac{C_{5V}}{m_N^2}(g^{\alpha \mu}q \cdot p-q^\alpha p^\mu)+C_{6V}g^{\alpha\mu}\bigg] \gamma_5
\nonumber\\
\Gamma^{\alpha \mu}_A(p,q)&=&\frac{C_{3A}}{m_N}(g^{\alpha\mu}\slashed q -q^\alpha \gamma^\mu)+\frac{C_{4A}}{m_N^2}(g^{\alpha \mu}q\cdot p_\Delta-q^\alpha p_\Delta^\mu)
    +C_{5A}g^{\alpha \mu}+\frac{C_{6A}}{m_N^2}q^\alpha q^\mu\,.
\end{eqnarray}
The \(C_{iV/A}\) form factors depend on \(q^2\) and are given by the following expressions~\cite{Hernandez:2007qq}:
\begin{equation}
\begin{split}
    C_{3V}(q^2)=\frac{2.13}{\big(1-\frac{q^2}{M_V^2}\big)^2}\times \frac{1}{1-\frac{q^2}{4M_V^2}}\,, \qquad C_{4V}(q^2)=\frac{-1.51}{\big(1-\frac{q^2}{M_V^2}\big)^2}\times \frac{1}{1-\frac{q^2}{4M_V^2}}\,, \\  C_{5V}(q^2)=\frac{0.48}{\big(1-\frac{q^2}{M_V^2}\big)^2}\times \frac{1}{1-\frac{q^2}{0.776 M_V^2}}\,, \qquad C_{6V}(q^2)=0 \, ;\qquad \qquad
    \end{split}
\end{equation}
\begin{equation}
\begin{split}
    C_{3A}(q^2)=0\,, \qquad C_{4A}(q^2)=-\frac{C_{5A}(q^2)}{4}\,,\qquad\qquad\qquad\\  C_{5A}(q^2)=\frac{1.2}{\big(1-\frac{q^2}{M_A^2}\big)^2}\times \frac{1}{1-\frac{q^2}{3M_A^2}}\,, \qquad C_{6A}(q^2)=C_{5A}(q^2)\frac{m_N^2}{m_\pi^2-q^2} \, ,
    \end{split}
\end{equation}
where \(M_{V/A}\) are the nucleon vector and axial masses respectively, reported in Tab.~\ref{tab:physq}.  

The computation of the isospin part of the semi-inclusive tensor is illustrated in detail with an example in Appendix \ref{app:Isospin}.

To make the RFG model more realistic, an energy shift is introduced, which phenomenologically accounts
for the nucleon binding energy and for final-state interaction effects. 
In our approach, the energy shift is a constant parameter, depending on the target and extracted by electron-nucleus scattering data. This procedure ensures that the position of the quasielastic peak is well reproduced. 
In the computation, this means that part of the transferred energy \(\omega\) is absorbed by the nucleus to eject a nucleon. Thus the elementary nucleon tensor is evaluated using an ``effective'' energy transfer \(\tilde \omega\), defined by
\begin{equation}
    \tilde \omega\equiv \omega-E_{shift}\,.
\end{equation}
Following the de Forest prescription \cite{DEFOREST1983232}, we evaluate the form factors using the exact \(\omega\)-value, while all the other terms in the hadronic tensor are affected by the energy shift.
The energy shift \(E_{shift}\) and the Fermi momentum \(p_F\)  represent the only two parameters of the RFG model, the latter being fixed by the target nucleus and fitted to the width of the quasielastic peak in $(e,e')$ data \cite{Maieron:2001it}.  
In Ref.~\cite{Belocchi:2024rfp} we studied the impact of different \(E_{shift}\) values on the semi-inclusive electron scattering cross section. Comparing with $^{12}$C$(e,e'p)$ data, we decided to use the value of \(E^{2p2h}_{shift}=40\) MeV, to make the theoretical results closer to the available data. Then, for the $^{12}$C nucleus, the following parameter values are chosen: $p_F=225$ MeV and $E_{shift}=20 $ MeV, with \(E^{2p2h}_{shift}=2E_{shift}\).

In the \(2p2h\) semi-inclusive hadronic tensor \eqref{eq:WmunuN}, a nine-dimensional integral must be evaluated to obtain the responses. 
Through appropriate manipulations, this integral can be reduced to a five-dimensional form by analytically exploiting the Dirac $\delta^4$. The procedure for this computational reduction is presented in detail in Appendix \ref{app:PS}.

\section{Results}
\label{sec:Results}

In this section we present the numerical results of the calculation illustrated above.
Before showing the semi-inclusive $2p2h$ cross sections, we validate our calculation by evaluating the inclusive weak \(2p2h\) response functions and comparing the results with the ones published in Ref.~\cite{RuizSimo:2016ikw}. The values of the parameters used in this work are listed in Tab.~\ref{tab:physq}.

\begin{table}[ht!]
\centering
\begin{tabular}{cccc}
\hline \hline
$\quad$ Physical quantities $\quad$		& $\quad$Values $\quad$&$\quad$ Masses$\quad$ & $\quad$ Values [GeV] $\quad$\\
\hline
\(f_{\pi NN}^2\)    & \( 4\pi\times 0.08\) & \(m_N\)        & \(0.939\) \\
\(g_A\)         &   \(1.26\) & \(m_\pi\)        & \(0.1395\) \\
\(f_\pi\)         &   \(93\) MeV  & \(m_\rho\)        & \(0.7758\)\\
\(f^*\)         &   \(2.14\) & \(M_\Delta\)        & \(1.232\)\\
\(\Lambda_\pi\)         & \(1.3\) GeV   & \(M_V\)        & \(0.84\)\\
 \(\Lambda_\Delta\)        &  \(1.15\) GeV  & \(M_A\)        & \(1.05\)\\
\hline \hline
\end{tabular}
\caption{Values of the couplings, cut-off parameters and masses adopted in this work, taken from Ref.~\cite{Hernandez:2007qq}.}
\label{tab:physq}
\end{table}

\subsection{Inclusive cross section}

\begin{figure}[ht!]
    \centering
    \includegraphics[width=1\linewidth]{ 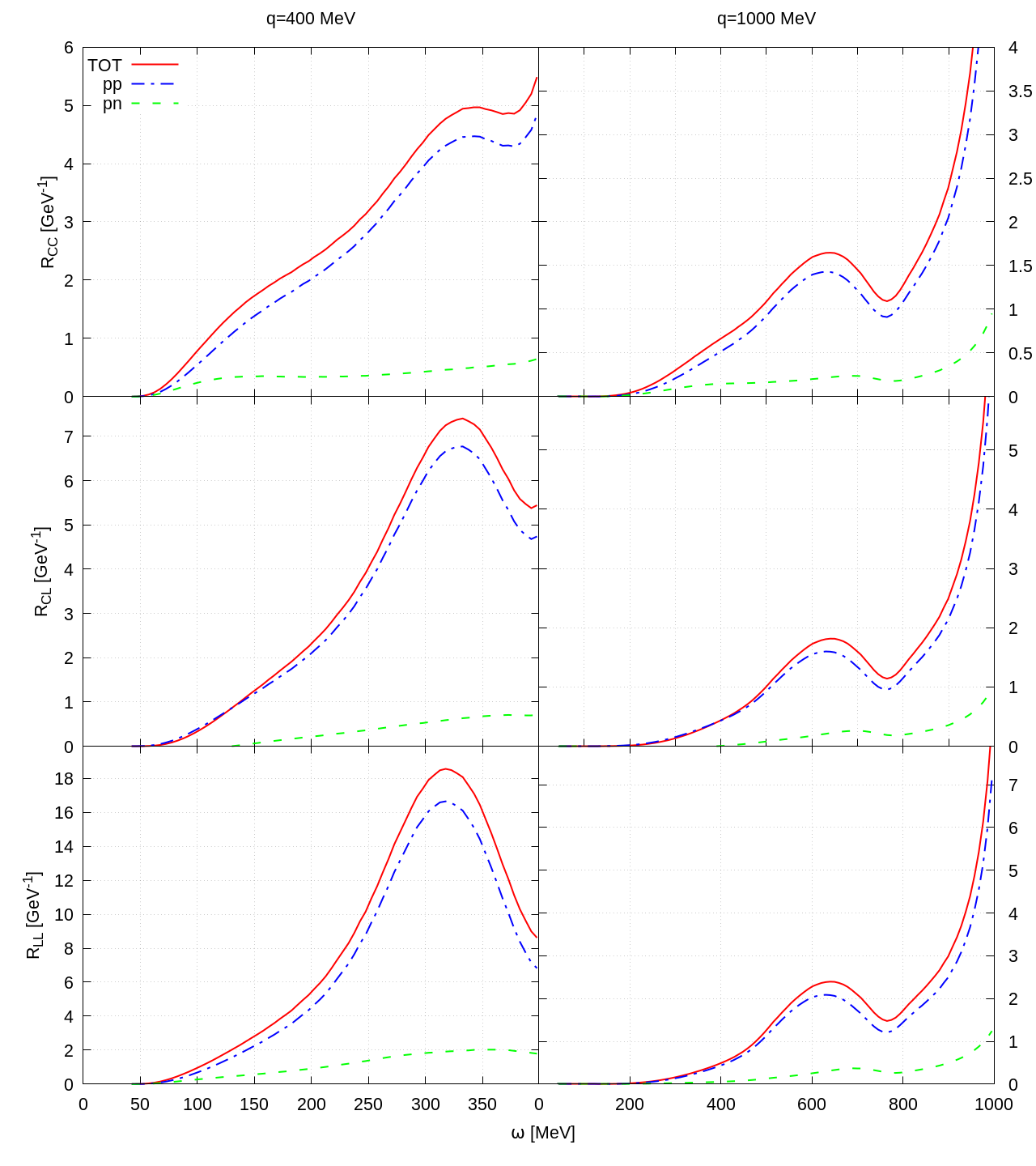}
    \caption{Weak \(2p2h\) $\nu_\mu$-$^{12}C$  inclusive longitudinal responses evaluated at \(|{\bf q}|=400\) and \(1000\) MeV. The solid lines represent the responses obtained including all possible two-nucleon pairs in the final states. The dot-dashed lines correspond to the \(pp\) final state contribution, while the dashed lines account for the \(pn\) final state. 
 The responses are calculated using only the dominant form factors \(C_{3V},\,C_{5A}\).}
    \label{fig:EWL-pp-pn}
\end{figure}

\begin{figure}[h!]
    \centering
\includegraphics[width=1\linewidth]{ 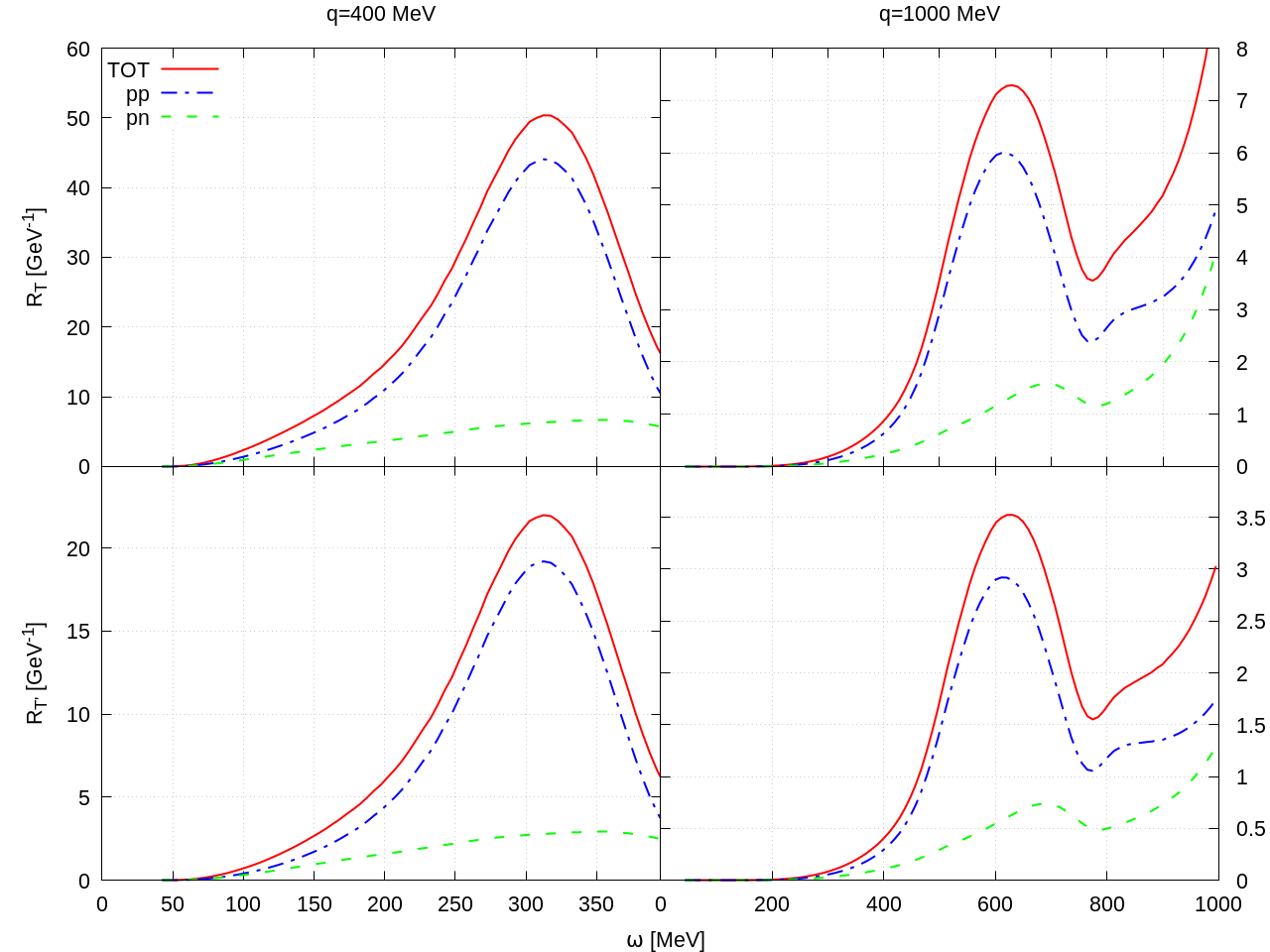}
    \caption{Weak \(2p2h\) $\nu_\mu$-$^{12}C$   inclusive transverse responses \(R_T\)(top) and \(R_{T'}\)(bottom) evaluated at \(|{\bf q}|=400\) and \(1000\) MeV as functions of \(\omega\). The solid lines represent the responses obtained including all possible two-nucleon pairs in the final states. The dot-dashed lines correspond to the \(pp\) final state contribution, while the dashed lines account for the \(pn\) final state. 
 The responses are calculated using only the dominant form factors \(C_{3V},\,C_{5A}\), consistently with Ref.~\cite{RuizSimo:2016ikw}.}
    \label{fig:EWT-pp-pn}
\end{figure}
\begin{figure}[h!]
    \centering
    \includegraphics[width=0.75\linewidth]{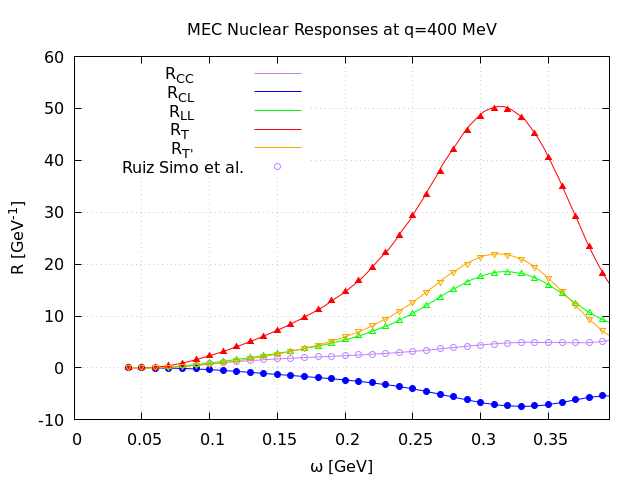}
    \caption{Weak nuclear inclusive responses evaluated at fixed \(|{\bf q}|=\,400\) MeV are compared with the computation of Ref. \cite{RuizSimo:2016rtu}. }
    \label{fig:comparison}
\end{figure}

The model used to calculate the inclusive $(\nu_\mu,\mu)$ cross section is exactly the same previously illustrated, except that now the cross section involves an extra integration over the unobserved outgoing nucleon variables. It can be written as a linear combination of five response functions:
    \begin{equation}
    \frac{\rm d^3 \sigma}{\rm d E_{k'}\rm d \Omega_{k'} }=\sigma_{0} \left(V_{CC}R_{CC}-2V_{CL}R_{CL}+V_{LL}R_{LL}+V_{T}R_{T} + 2V_{T'}R_{T'} \right)
\,,
    \label{xsec-incl}
\end{equation}
where each response $R_K(\omega,|\mathbf{q}|)$ depends on only two variables and follows the same definitions reported in Appendix~\ref{app:resp}, Eqs.~(\ref{eq:RCC}-\ref{eq:RTp}), but involving the components of the inclusive nuclear tensor $ W^{\mu \nu}_{A \,2p2h}$. This is connected to the semi-inclusive one by
\begin{equation}
    W^{\mu \nu}_{A \,2p2h}=\int W^{\mu \nu}_{A(N) \,2p2h}({\bf p_1})\,\ud {\bf p_1}\,,
\end{equation}
where the sum over every allowed isospin state is performed.

In Figs.~\ref{fig:EWL-pp-pn} and \ref{fig:EWT-pp-pn}, the  \(2p2h\) responses for $\nu_\mu$-$^{12}C$ inclusive scattering are displayed as functions of the transferred energy \(\omega\), at given momentum transfer \(q\), together with the separate contributions from $pp$ and $pn$ pairs in the final state. The proton-proton channel emerges as the dominant one, being the proton-neutron contribution about a sixth of the former. The two contributions show a different shape: the \(pp\) is peaked at lower \(\omega\) values, while the \(pn\) presents a wider bump, with the peak located at higher \(\omega\).  In Fig.~\ref{fig:comparison} our results are  compared with the responses presented in Ref.~\cite{RuizSimo:2016ikw}, showing perfect agreement.
In these results the \(\Delta\)  interaction vertex is described  using only the dominant form factors, to be consistent with Ref.~\cite{RuizSimo:2016ikw}.
Note, however, that in the other results  presented in this work all the $\Delta$ form factors appearing in the resonance interaction vertex are included. The effect of the subdominant form factors is not negligible, and it is particularly important in the CC, CL and LL responses, as shown in Ref.~\cite{Belocchi:2025lqp}.

\subsection{Semi-inclusive cross sections}

When considering a semi-inclusive process such as $(\nu_\mu,\mu N)$, the cross section is evaluated as a function of the three-momentum carried by one of the nucleons in the final state. Usually, this is chosen to be a proton, because it is the one that is easier to detect experimentally. Note that, unlike in the electromagnetic case, in CC neutrino-nucleus scattering a proton is always emitted in the \(2p2h\) channel, not considering further interactions experienced by the outgoing nucleons. 

In order to study how the neutrino-nucleus cross section behaves at different final proton kinematics, the sixth differential cross section given in Eq.~\eqref{eq:xsec-semi1} is evaluated, fixing the incident neutrino energy \(E_k\equiv E_{\nu_\mu}\), the scattering angle \(\theta_{k'}\equiv\theta_\mu\) and the transferred energy \(\omega\).
The reference considered is the q-system, so that each angle is relative to the \({\bf q}\) direction.

\begin{figure}[ht!]
    \centering
    \includegraphics[width=1\linewidth]{ 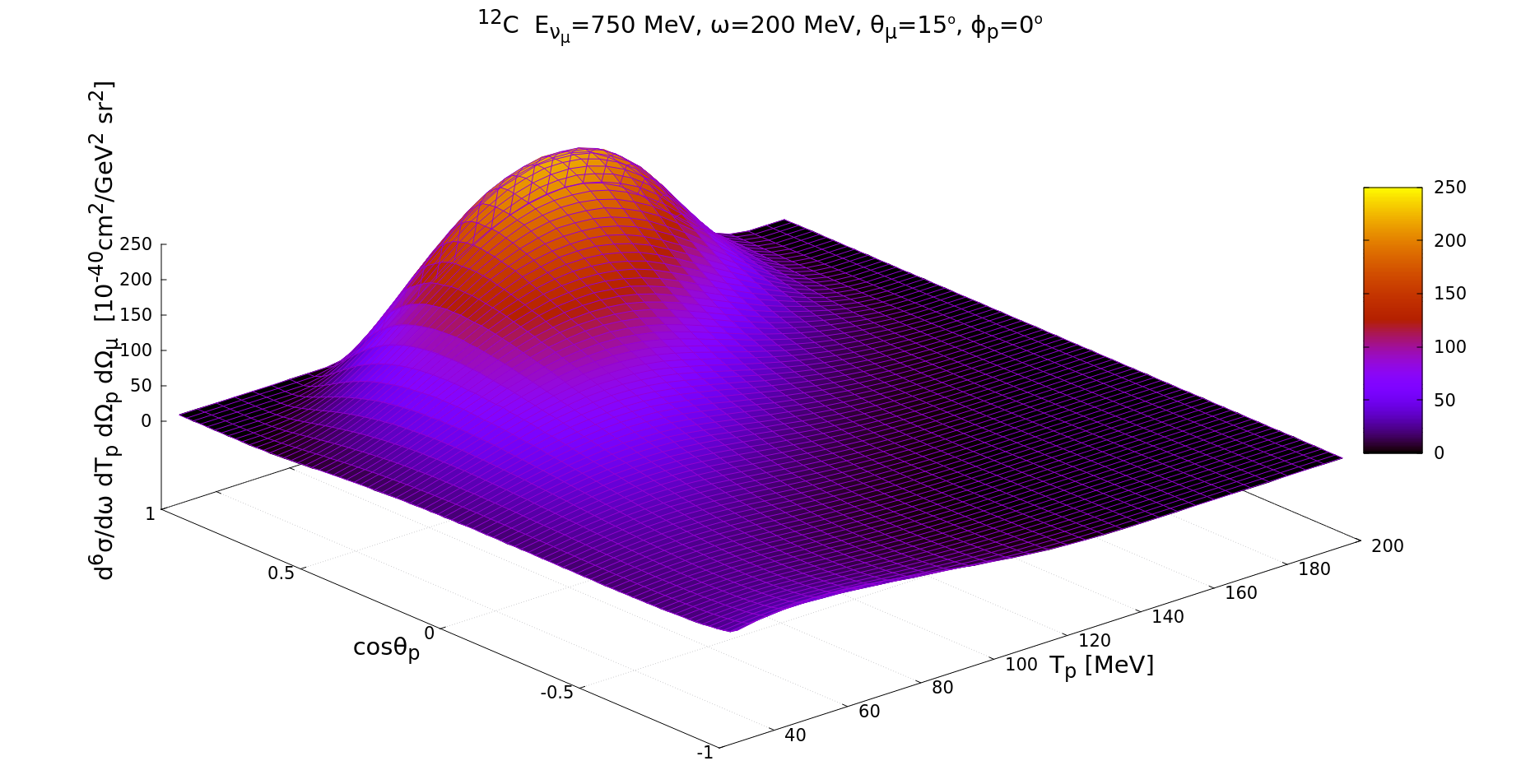}\\
    \includegraphics[width=1\linewidth]{ 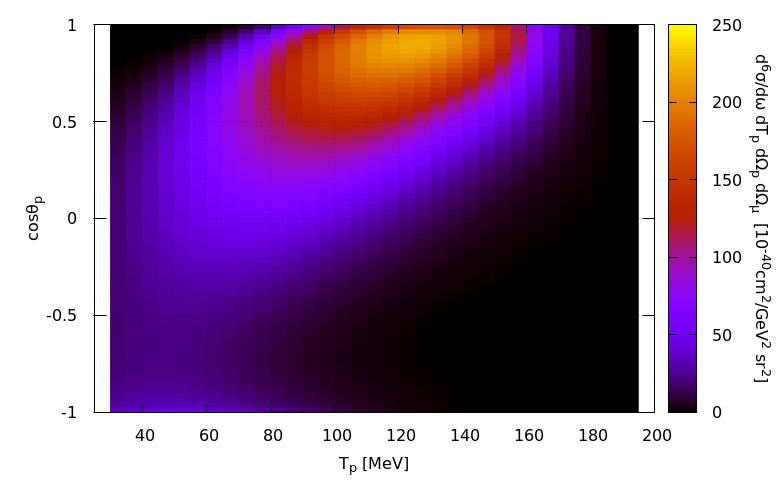}
    \caption{Semi-inclusive $\nu_\mu$-$^{12}C$ sixth differential cross section for the \(2p2h\) channel, computed at incident muonic neutrino energy \(E_{ \nu_\mu}=750\) MeV and transferred energy \(\omega=200\) MeV, displayed as a function of the polar angle \(\theta_p\) and kinetic energy \(T_p\) of the final proton. The scattering angle \(\theta_\mu=\)15° is fixed, as well the azimuthal final proton angle \(\phi_p=0\)°. The corresponding momentum transfer is \({\bf |q|}\simeq268\) MeV. The bottom panel is the two-dimension projection of the upper one, with the colors representing the z-axis.  }
    \label{fig:3D-EW15}
\end{figure}
        
In the three-dimensional plot of 
Fig. \ref{fig:3D-EW15} the cross section, computed at \(E_{\nu_\mu}=750\) MeV, \(\omega=200\) MeV and for scattering angle \(\theta_\mu=15^\circ\),
is displayed as a function of the proton polar angle and kinetic energy.
Only this figure representing the full-differential cross section is included in the main text, to improve readability. Other figures, starting from Fig.~\ref{fig:3D-30-60deg}, which is the equivalent of Fig.~\ref{fig:3D-EW15} but for scattering angles \(\theta_{\mu}=30^\circ\) and \(60^\circ\), are reported in Appendix~\ref{app:3D} but discussed here. In all presented kinematics, the differential cross section is peaked nearby \(\theta_p=0^\circ\), that is the so-called ``parallel kinematics''. The peak is well defined, centered at higher \(T_p\) values when the polar angle of the ejected proton is small. The strength of the cross section is reduced at higher \(\theta_p\) values, and consequently the peak moves to lower proton kinetic energies. This behaviour is clearly understood: the three-momentum conservation at fixed \(\omega\) forces the final proton to be ``forward" respect to the three-momentum transfer \({\bf q}\). In this condition, the energy of the on-shell final proton can be higher. Conversely, at higher proton angle the proton momentum is reduced, and so is the related energy.
In this kinematical region, as it is possible to see in Fig. \ref{fig:3D-60deg_study}, in the backward region the cross section is strongly reduced. This is also due to the negative interplay between the pion's and resonance's contributions.

The kinematics analyzed in Fig.~\ref{fig:3D-EW15} and Figs.~\ref{fig:3D-30-60deg}-\ref{fig:3D-15pp_pn} are chosen in order to compare our results with those of Refs.~\cite{Niewczas:2023gii, VanCuyck:2016fab}. The same kinematics are also investigated for more exclusive \(2p2h\) cross sections in Ref.~\cite{sym16020247}. In fact in this kinematical region the \(2p2h\) channel gives an important contribution to the cross section, especially at small scattering angles. For instance, at \(\theta_\mu=15\)° the cross section is dominated by \(2p2h\), being the QE peaked at \(\omega_{\rm QE}=Q^2/2m_N\simeq37\) MeV - including \(E_{shift}\) -, while the pion production, starting when the energy transfer is higher than the pion mass, is still small.

In Fig.~\ref{fig:3D-60deg_study},  at \(\theta_\mu=15\)°, the  contributions from the \(\Delta\), the purely pionic,  and the interference \(\pi-\Delta\) parts of the MEC to the total cross section are displayed. The \(\Delta\) current represents the dominant contribution of the MEC, providing more than a half of the total strength. This contribution shows a defined bump centered in the proton parallel kinematic, at \(T_p\simeq 150\) MeV, that rapidly decreases at higher polar angles and higher kinetic energy. Conversely, reducing \(T_p\) the cross section decreases softly. The pionic MEC part presents a similar behaviour but the peak, situated at the same position of the \(\Delta\) contribution, is more localized, with a steep behaviour moving away from the peak. 

The interference contribution, although subdominant, is not negligible. Its sign and magnitude exhibit a strong dependence on the underlying kinematics. In the present configuration, the interference term undergoes a sign change, while its strength displays a broad bump in the \(T_p\) range, localized at very forward angles. A rapid suppression is observed toward transverse proton emission, {\it{id est}} at \(\theta_p\simeq 90\)°, where the contribution becomes small and negative. In contrast, backward emission does not receive any appreciable contribution from pion–resonance interference. The \(\pi-\Delta\) cross section exhibits a similar proton-angular dependence at \(\theta_\mu= 30\)°, characterized by a more localized bump within the \(T_p\) range. Conversely, at \(\theta_\mu= 60\)° the contribution remains consistently negative.

In Fig.~\ref{fig:3D-15pp_pn}, the contributions from the different final isospin channels are evaluated separately for \(\theta_\mu=15\)°, to facilitate comparison with the results presented in Ref.~\cite{Niewczas:2023gii}. The \(pp\) channel exhibits a broader bump and dominates over the \(pn\) channel, which is more localized at higher \(T_p\) values, with a peaked distribution.
In our results the ratio \(pp/pn\) is \(\simeq 4\). Although this quantity depends on the final proton kinematics, and a three-dimensional plot is not ideal for quantifying it, nonetheless it is slightly different from the ratio presented in Ref.~\cite{Niewczas:2023gii}, where a higher \(pp/pn\) result is reported. It is important to note that the model in \cite{Niewczas:2023gii} incorporates a non-relativistic reduction of the MEC but also includes short-range correlations, which could partially account for the discrepancy.
Our result for the isospin \(pp/pn\) ratio is more consistent with the findings in Ref.~\cite{sym16020247}, although in this reference different exclusive observables were calculated, such as the eightfold differential cross section at fixed \(\omega,\,{\bf q},\,T_p\), with explicit dependence on the final nucleons' angles in the laboratory frame, a reference system discussed in the following section.

\begin{figure}
    \centering
    \includegraphics[width=1\linewidth]{ 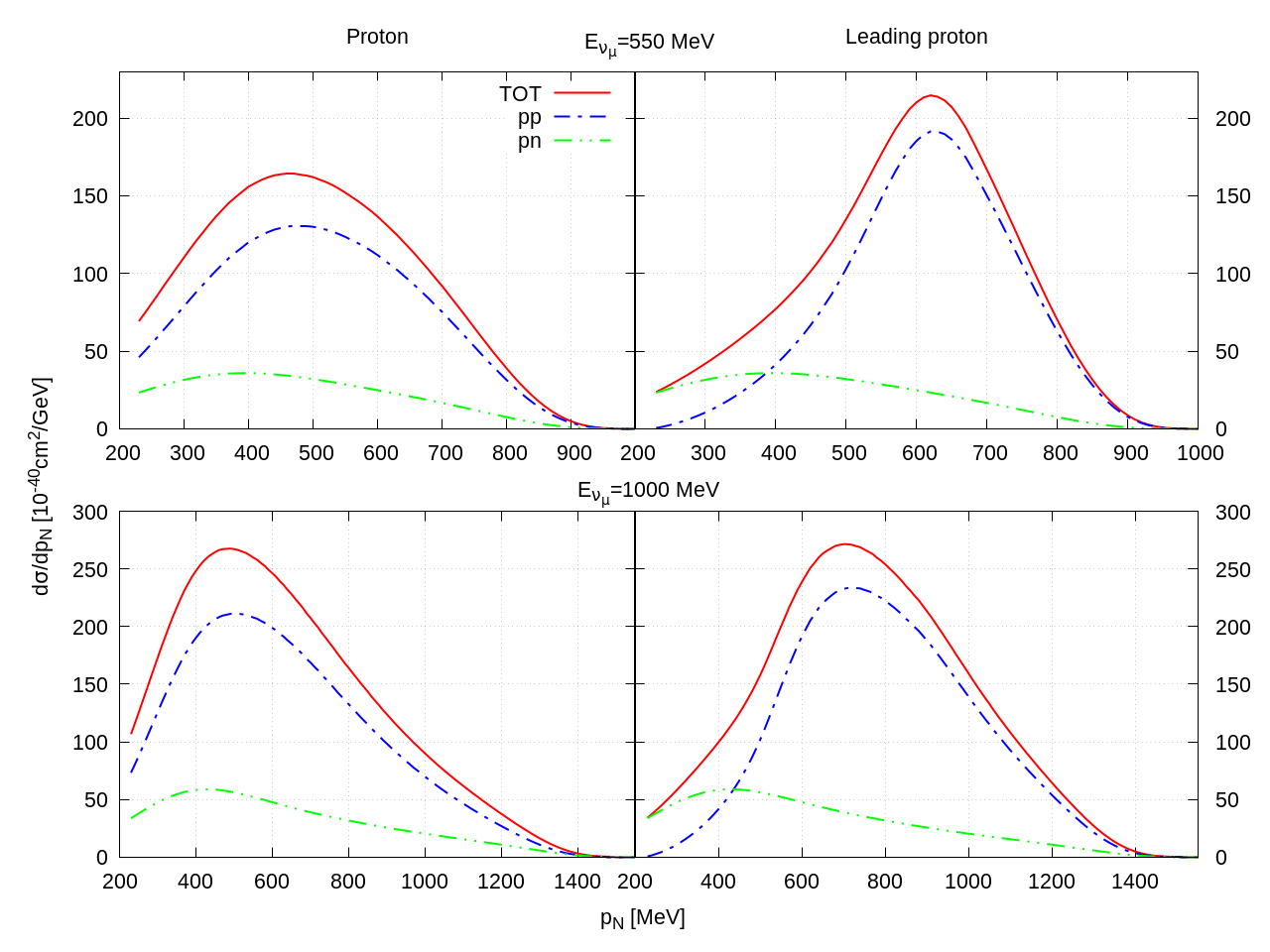}
    \caption{Weak $\nu_\mu$-$^{12}C$ CC semi-inclusive cross section evaluated at incident neutrino energy \(E_k=\)550 (top), 1000 (bottom) MeV. In the left panels the cross section is computed varying the final proton momentum, while in the right ones the leading proton momentum is used.}
    \label{fig:EW-leadproton}
\end{figure}

In Fig.~\ref{fig:EW-leadproton}, the differential semi-inclusive cross section 
\begin{equation}
    \frac{\rm d \sigma}{\ud p_N}=\int \ud \omega \ud \cos \theta_{k'} 2 \pi \sigma_{0}\frac{\ud \mathcal{F}_N^2}{\ud p_N}
    \label{eq:EWdsdpN}
\end{equation}
is computed fixing the incident neutrino energy. It is important to note that, as a consequence of the integration over all possible directions of the final proton, only the five  responses appearing in the first row of Eq.~\eqref{eq:F2} give a non-vanishing contribution to the cross section \eqref{eq:EWdsdpN}.
The differential semi-inclusive cross section \eqref{eq:EWdsdpN} can also be represented as a function of the momentum of the final ``leading'' proton,  defined as the proton carrying the highest momentum in a multi-nucleon knockout event. In the context of \(2p2h\) scattering process, the leading proton corresponds to the nucleon with the highest energy in a \(pp\) final state pair, while in the \(pn\) channel it is the only proton present in the final state.

The effect of the leading proton definition is examined in Fig.~\ref{fig:EW-leadproton} by comparing the differential cross section \eqref{eq:EWdsdpN} as a function of the proton momentum (left panels)  and of the leading proton momentum (right panels). As shown, the contribution from the \(pn\) final state is identical in both approaches, while the \(pp\) final-state channel is significantly affected. This difference is expected: using the leading proton variable, when two protons are present in the final state, the computed cross section is determined by the most energetic proton. Consequently, the cross section associated with small final-proton momenta in the \(pp\) channel is significantly suppressed. 
This phenomenon is more pronounced at low incident neutrino energy, caused by the peak shape and position: it is broad and centered at lower final-proton momentum. In this region, the leading proton definition has a more significant impact, enhancing the peak, which shifts to higher proton momentum values, while reducing the tails. In contrast, at higher energies both approaches yield similar shapes, although the peak is shifted towards higher proton momentum values when the cross section is displayed versus the leading proton momentum.

\begin{figure}
    \centering
    \includegraphics[width=1\linewidth]{ 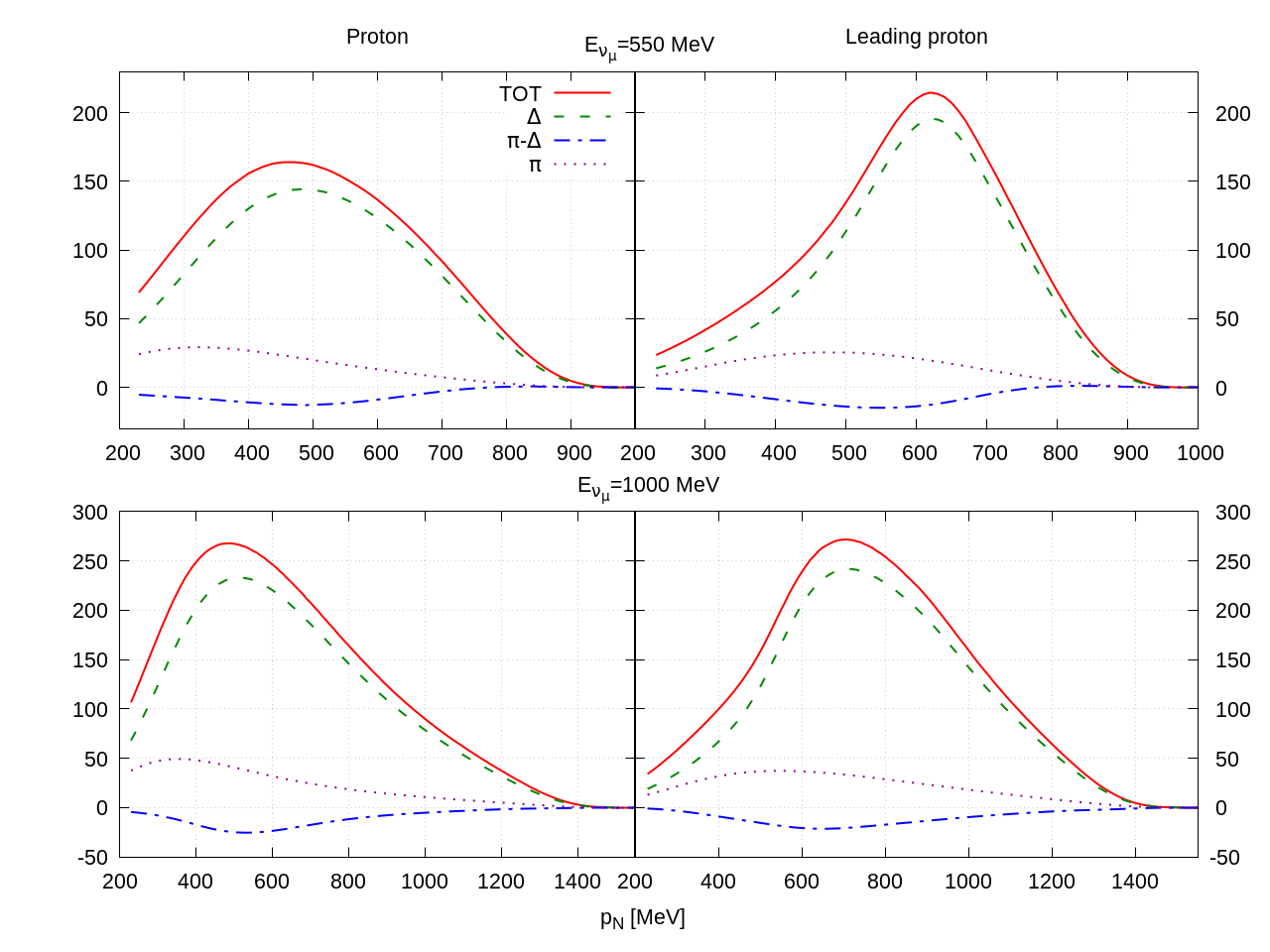}
    \caption{Weak $\nu_\mu$-$^{12}C$ CC semi-inclusive \(\Delta,\,\pi,\,\pi-\Delta\) contributions to the cross-section evaluated at incident neutrino energy \(E_{\nu_\mu}=\)550 (top), 1000 (bottom) MeV. In the left panels the cross section is computed varying the final proton momentum, while in the right ones the leading proton momentum is used.}
    \label{fig:EW-leadproton-sep}
\end{figure}

In Fig. \ref{fig:EW-leadproton-sep} the  contributions from the \(\Delta\), the purely pionic, and the interference \(\pi-\Delta\) parts of the MEC to the total cross section at the same kinematics are displayed. The \(\Delta\) current represents the very dominant contribution of the MEC, providing around the 80\% of the total strength. The pion component is subdominant, while the \(\pi-\Delta\) interference is consistently the smallest, and negative contribution. 

To make contact with experimental data, the flux-folded differential semi-inclusive cross sections are also calculated, using the T2K semi-inclusive $1\mu$CC$0\pi Np$ measurement \cite{T2K:2018rnz} as a reference. Specifically, the experimental kinematic cuts are applied in the calculation, which impose significant constraints on the phase space of the final-state particles. These cuts include restrictions on the outgoing muon scattering angle, as well as on the polar angle and on the momentum of the final proton in the laboratory frame. The usage of quantities related to the laboratory frame requires a specific rotation procedure, illustrated in Appendix~\ref{App3}.

\begin{figure}
    \centering
    \includegraphics[width=0.5\linewidth]{ 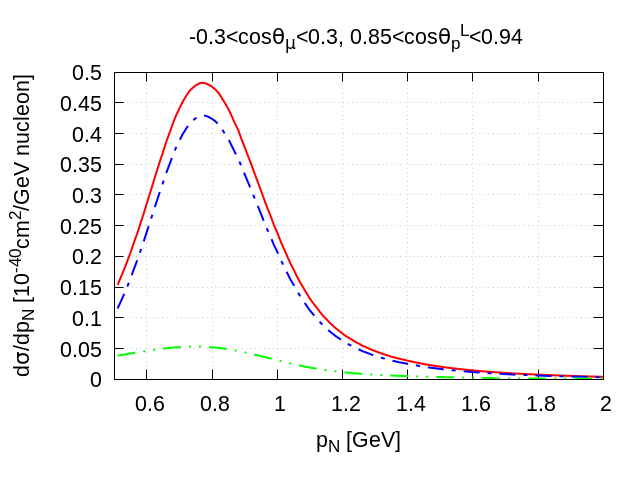}\includegraphics[width=0.5\linewidth]{ 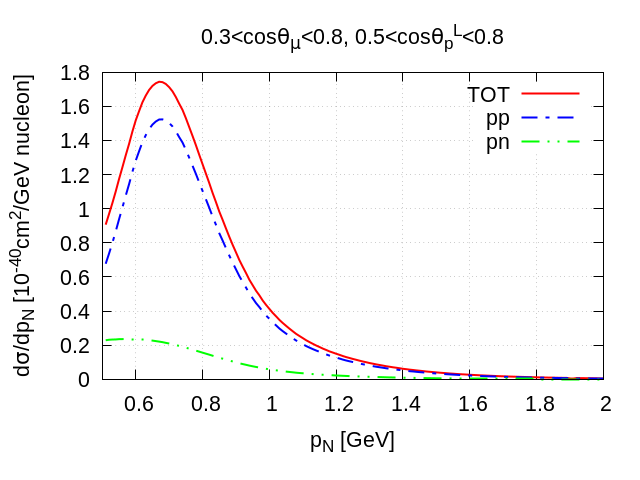}\\
    \includegraphics[width=0.5\linewidth]{ 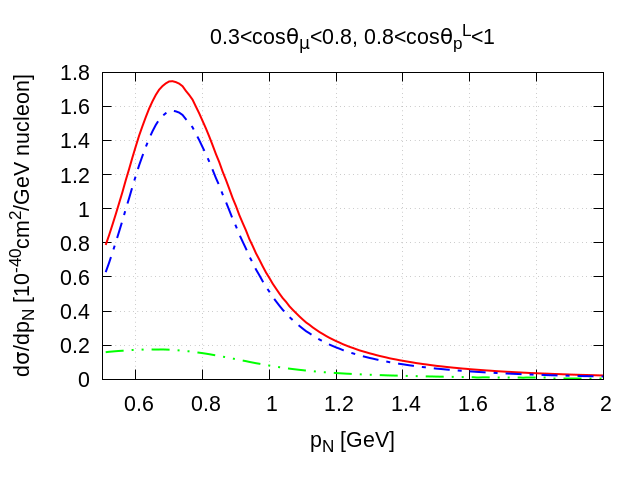}\includegraphics[width=0.5\linewidth]{ 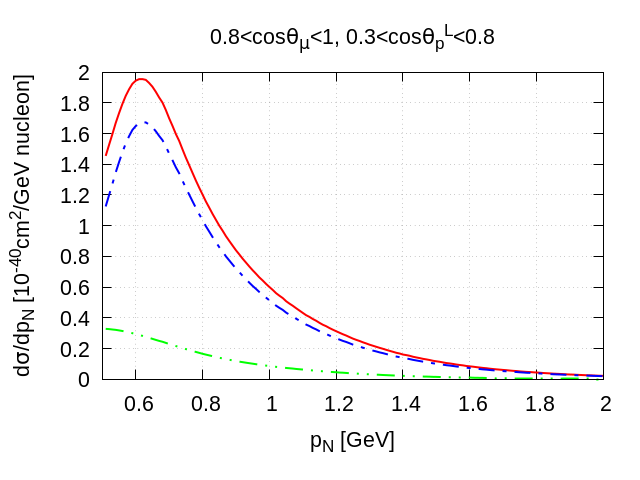}
    \caption{\(2p2h\) contribution to the single-differential \(1\mu\)CC\(0\pi Np\) cross section averaged over the incident T2K ND280 muon neutrino flux. Results are shown as functions of the final leading proton momentum \(p_N\). The T2K \cite{T2K:2018rnz} kinematical cuts on the muon scattering angle \(\theta_\mu\) and  laboratory-frame final proton polar angle \(\theta_p^L\), indicated in the titles of the figures, are applied. 
    The results are normalized for single active nucleon taking into account  the mineral oil target CH. Thus, the cross sections are obtained for carbon \(^{12}\)C and then divided by thirteen. The separate contributions of proton-proton \(pp\) and proton-neutron \(pn\) pairs in the final state are also displayed.}
    \label{fig:dsigma-dpN_T2K}
\end{figure}
\begin{figure}
    \centering
    \includegraphics[width=0.495\linewidth]{ 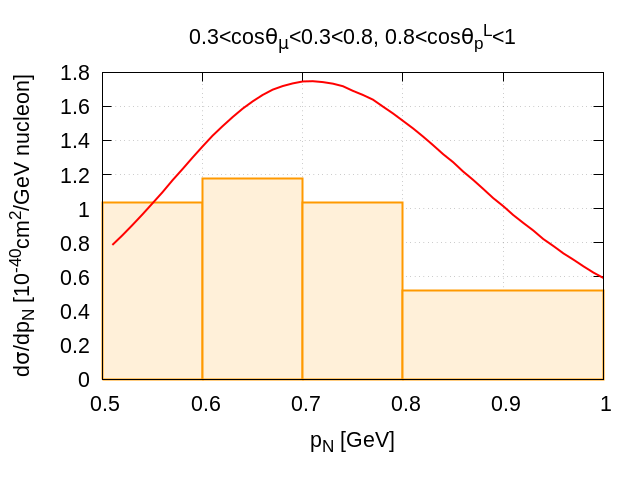} \includegraphics[width=0.495\linewidth]{ 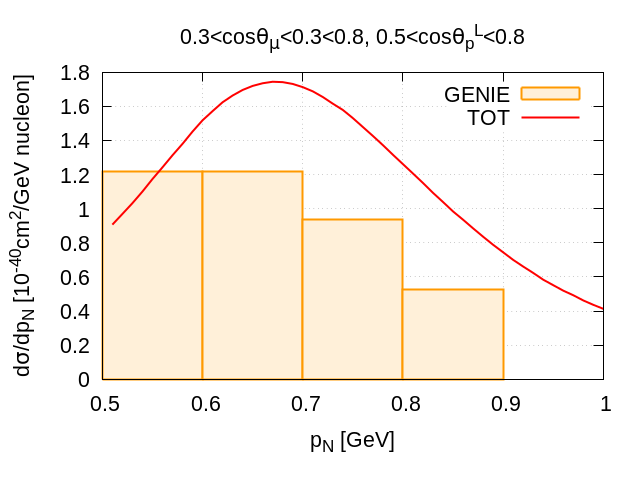}\\
    \caption{Differential  \(2p2h\) contribution to the \(1\mu\)CC\(0\pi Np\) cross section averaged over the incident T2K ND280 muon neutrino flux with T2K \cite{T2K:2018rnz} kinematical cuts compared to ``GENIE'' \cite{Dolan:2019bxf} theoretical predictions. The results are normalized for single active nucleon taking into account  the mineral oil target CH.}
    \label{fig:EW-Genie_comp}
\end{figure}

In Fig.~\ref{fig:dsigma-dpN_T2K}, we display the \(2p2h\) contribution to the \(1\mu\)CC\(0\pi Np\)  differential cross section, integrated over the T2K muonic neutrino flux, as a function of the leading ejected proton momentum \(p_N\). The separate contributions of proton-proton \(pp\) and proton-neutron \(pn\) pairs in the final state are also displayed. This result represents the first evaluation of the \(2p2h\) impact in this kind of process obtained from a fully microscopic, relativistic, {\it\underline{semi-inclusive}} computation. In the literature, the only  theoretical prediction available for this cross section is the one obtained with the implementation in GENIE of the SuSAv2-MEC model \cite{Dolan:2019bxf}, which incorporates the \(2p2h\) {\it\underline{inclusive}} model outlined in Ref.~\cite{RuizSimo:2016rtu} and suffers from the already mentioned inconsistencies.

Finally, in Fig.~\ref{fig:EW-Genie_comp} we compare our results with the Monte Carlo predictions of Ref.~\cite{Dolan:2019bxf}. The two results seem to be compatible, both in terms of shape and order of magnitude; however, systematic discrepancies are observed. Our results consistently show a higher strength compared to the GENIE predictions, with a peak occurring at a higher final leading proton momentum. 
In order to understand the origin of these discrepancies, it should be stressed that the SuSAv2-MEC implementation in GENIE \cite{Dolan:2019bxf} is based on an inclusive theoretical model and requires certain assumptions in order to ``extract'' semi-inclusive predictions from the inclusive results. This process in principle is not possible since the semi-inclusive cross section involves kinematic variables which are integrated over in the inclusive one. On the contrary, our calculation deals correctly with all the kinematic variables. As a result, differences between our model and the Monte Carlo outcomes are expected.
On the other hand, the SuSAv2-MEC Monte Carlo implementation in GENIE benefits from the incorporation of additional nuclear effects, absent in our calculation, such as final state interactions (FSI), which are modeled in GENIE using a semi-classical cascade approach.  In general, FSI tend to broaden and shift the differential cross section towards lower final proton energy and momentum. When considering only protons with momentum \(p_N>500\) MeV, this effect results in a reduction in the cross section strength. The cascade model accounts for multiple re-scattering processes of the ejected nucleon with other nucleons in the nucleus, which also influences the angular distribution of the final proton. In this context, kinematical angular cuts may further alter both the shape and strength of the cross section. 

A comprehensive comparison with the T2K data in the kinematics above discussed is presented in Ref.~\cite{Franco-Patino:2022tvv}, Fig.~4, where the results of a semi-inclusive calculation in the quasi-elastic channel are presented. In this reference, the sensitivity to the model used to describe the FSI between the ejected proton and the residual nucleus is investigated and shown to be very relevant for semi-inclusive predictions. However, the comparison of theoretical predictions with data is strongly influenced by the contribution of \(2p2h\) events to the experimental signal. Thus, a reliable evaluation of this contribution is also required for a correct interpretation of the semi-inclusive data.

\section{Conclusions}
\label{sec:Concl}

Motivated by the promising results presented in \cite{Belocchi:2024rfp} for the $2p2h$  contribution to  the electromagnetic $(e,e'p)$ process, in this work we have extended the calculation to the weak sector and made  predictions for neutrino-carbon semi-inclusive $(\nu_\mu,\mu^-p)$ observables. 

The weak model has been first validated by comparing with the inclusive results  of Ref.~\cite{RuizSimo:2016rtu}, which employs the same  nuclear model and current operators. Moreover, we have analyzed the individual  contributions from different isospin channels. We have found that the $pp$ emission channel dominates over the $pn$ one, and that the two cross sections exhibit different dependencies on the transferred energy $\omega$.

For the semi-inclusive channel, we have presented  sixfold differential neutrino-carbon  cross sections as functions of one ejected proton momentum and angle, under specific kinematics corresponding to available T2K data \cite{T2K:2018rnz}. We have carried out a systematic study across the final proton phase-space at different kinematics, evaluating the  \(pp\) and \(pn\) emission channels. The cross sections have also been decomposed into  contributions from the \(\pi,\,\pi-\Delta\) interference, and \(\Delta\) diagrams. Noticeably, we have found that the \(\pi-\Delta\) interference can be either constructive or destructive, depending on the kinematics, and is far from negligible. As in the inclusive case, the \(\Delta\) contribution is dominant.

To facilitate comparisons with semi-inclusive experimental data — typically expressed in terms of the momentum of the "leading proton" — we have also computed the differential cross section \( \ud \sigma/\ud p_N\) using both the detected proton momentum and the leading proton momentum (namely, the highest momentum among the ejected nucleon pair), assessing  the impact of different definitions on the observable.

Then, we have presented theoretical predictions for the \(2p2h\) contribution to the semi-inclusive  \(1\mu\)CC\(0\pi Np\)  cross section, folded over the incident T2K ND280 muon neutrino flux and shown as a  function of the  leading proton momentum \(p_N\). The T2K kinematical cuts have been applied to estimate the impact of the \(2p2h\) contribution on the observed data. 

Finally, we have performed a direct comparison with the outcome of the GENIE simulation \cite{Dolan:2019bxf} for these data. The latter is performed in the  framework of Meson Exchange Currents on a RFG basis, but
uses inclusive rather than semi-inclusive results, resorting to some necessary approximations to predict semi-inclusive observables.
Several differences have emerged from this comparison, concerning the strength of the two-particle two-hole contribution. This suggests that event generators might require the adoption of more exclusive models to properly simulate semi-inclusive processes. Our approach is a good candidate for this improvement, offering a good balance between accuracy and computational efficiency. In fact, despite the very high number of terms appearing in the computation, the computing time is reasonable:
 for a given four-momentum transfer, the differential cross section can be computed with ~1\% numerical uncertainty in about one minute of CPU on an average-performance computer—comparable to the inclusive case.

A related model for the exclusive \(2p2h\) was recently developed in Ref.~\cite{sym16020247}, where the same RFG-based MEC formalism adopted in this work is enriched by introducing the effective mass for the on-shell nucleons, evaluated in a mean field model.
 Although a direct numerical comparison is not feasible -due to differences in implementation or available observables- the two approaches appear to be in qualitative agreement.
 
The calculation presented in this work paves the way for a more accurate description of the \(2p2h\) interaction process driven by meson-exchange currents.  The present model is based on the simple Relativistic Fermi Gas (RFG) in the infinite nuclear matter approximation, and it still lacks important nuclear effects, such as nucleon-nucleon correlations and final state interactions (FSI). 
While the complexity of the calculation necessitates certain approximations, accounting for medium effects remains a potential area for extensions. 

Future developments could include the adoption of more realistic nuclear frameworks, such as RFG-based Hartree-Fock or relativistic mean-field models. Additionally, the impact of incorporating the $\rho$ meson into the MEC formalism -though expected to be smaller than that of the pion- is under investigation.

It is important to emphasize that multi-nucleon excitations alone cannot fully explain the experimental data, as the concurrent presence of quasi-elastic and other interaction channels complicates direct comparison. Nonetheless, this work marks a major step towards a detailed and consistent modeling of the semi-inclusive $2p2h$ channel.

Overall, this calculation constitutes a novel contribution to neutrino–nucleus interaction studies. It provides a robust framework for generating exclusive predictions in the 
$2p2h$ channel, supporting the analysis of current and future high-precision measurements from accelerator-based neutrino experiments.

\begin{acknowledgements}
We thank Luis Alvarez Ruso, José Enrique Amaro and Paloma Casalé for useful discussions.
This work was partially supported by the Project NUCSYS funded by INFN; by the Grant BARM-RILO-24-01 funded by University of Turin; by the
“Planes Complementarios de I+D+i” program (Grant
ASFAE/2022/022) by MICIU with funding from the
European Union NextGenerationEU and Generalitat
Valenciana; by grant PID2023-147458NB-C21 funded by MCIN/AEI/ 10.13039/501100011033 and by the European Union.
\end{acknowledgements}

\appendix

\renewcommand{\thefigure}{\thesection.\arabic{figure}}
\setcounter{figure}{0}

\section{Leptonic factors and response functions}
\label{app:resp}
The 
leptonic factors appearing in the definition \eqref{eq:F2} of \(\mathcal{F}^2_{N}\)  are 
\begin{align}
V_{CC}&\equiv \frac{\tilde L^{00}}{\nu_0}=1-\delta^2\tan^2\tilde{\theta}/2\\
    V_{CL}&\equiv \frac{\tilde L^{03}+\tilde L^{30}}{2\nu_0}=\frac{\lambda}{\kappa}+\frac{\delta^2}{\rho'}\tan^2\tilde{\theta}/2 \\
    V_{LL}&\equiv \frac{\tilde L^{33}}{\nu_0}=\Big(\frac{\lambda}{\kappa}\Big)^2+\delta^2\tan^2\tilde{\theta}/2\Big(1+\frac{\lambda}{\kappa}\frac{2}{\rho'}+\rho\delta^2 \Big)\\
    V_T&\equiv \frac{\tilde L^{11}+\tilde L^{22}}{2\nu_0}=\frac{\rho}{2}+\tan^2{\tilde{\theta}/2}-\delta^2\tan^2{\tilde{\theta}/2}\Big(\frac{\lambda}{\kappa}\frac{1}{\rho'}+\frac{1}{2}\rho \delta^2\Big)\\
    V_{T'}&\equiv i\frac{\tilde L^{12}-\tilde L^{21}}{2\nu_0}=\frac{\tan^2{\tilde{\theta}/2}}{\rho'}\bigg(1-\frac{\lambda}{\kappa}\rho'\delta^2\bigg)>0\\
V_{TT}  &\equiv \frac{\tilde{L}^{11}-\tilde{L}^{22}}{2\nu_0}=\frac{\rho}{2}-\delta^2 \tan^2{\tilde{\theta}/2}\bigg(1+\frac{1}{\rho'}\frac{\lambda}{\kappa}+\frac{\rho\delta^2}{2} \bigg)>0\\
V_{CT}  &\equiv \frac{\tilde{L}^{01}+\tilde{L}^{10}}{2\nu_0}=\frac{1}{\rho'}\sqrt{2\frac{\tan^2{\tilde{\theta}/2}}{\rho}V_{TT}}\\
V_{LT}  &\equiv \frac{\tilde{L}^{31}+\tilde{L}^{13}}{2\nu_0}=\bigg(\frac{\lambda}{\kappa}+\rho'\rho\delta^2 \bigg)V_{CT}\\
V_{C\bar{T}}  &\equiv \frac{\tilde{L}^{02}+\tilde{L}^{20}}{2\nu_0}=0\\
V_{L\bar{T}}  &\equiv \frac{\tilde{L}^{32}+\tilde{L}^{23}}{2\nu_0}=0\\
V_{C\bar{T}'}  &\equiv i\frac{\tilde{L}^{02}-\tilde{L}^{20}}{2\nu_0}=\rho'{V_{CT}}\\
V_{L\bar{T}'}  &\equiv i\frac{\tilde{L}^{32}-\tilde{L}^{23}}{2\nu_0}=\frac{\lambda}{\kappa}V_{C\bar{T}'}\\
V_{CT'}  &\equiv i\frac{\tilde{L}^{01}-\tilde{L}^{10}}{2\nu_0}=0\\
V_{LT'}  &\equiv i\frac{\tilde{L}^{31}-\tilde{L}^{13}}{2\nu_0}=0\,.
\end{align}
The \(T\) leptonic factors related to the \(y\) tensor components vanish due to the choice of the conventional scattering plane, corresponding to the \(xz\) plane, and so do the \(T'\) terms related to \(x\). Note that there are different conventions describing the semi-inclusive leptonic factors and responses, such as that illustrated in Ref. \cite{Franco-Patino:2020ewa}: what matters is the coherent definition of the corresponding nuclear responses. In particular, in our conventions \(V_{TT}\) is chosen to be positive, as the other factors.

The nuclear responses are listed below:
\begin{align}
R_{CC}^{(N)}&\equiv W_A^{00} \label{eq:RCC}\\
    R_{CL}^{(N)}&\equiv\frac{1}{2}\big( W_A^{03}+W_A^{30}\big) \label{eq:RCL}\\
    R_{LL}^{(N)}&\equiv W_A^{33} \label{eq:RLL}\\
    R_T^{(N)}&\equiv W_A^{11}+W_A^{22} \label{eq:RT}\\
    R_{T'}^{(N)}&\equiv\frac{1}{2} \mathrm{Im}\big(W_A^{12}-W_A^{21} \big) \label{eq:RTp}\\
R_{TT}^{(N)}&\equiv W_{A(N)}^{11}-W_{A(N)}^{22}\\
R_{CT}^{(N)}&\equiv W_{A(N)}^{01}+W_{A(N)}^{10}\\
R_{LT}^{(N)}&\equiv W_{A(N)}^{31}+W_{A(N)}^{13}\\
R_{C\bar{T}'}^{(N)}&\equiv \mathrm{Im}\big(W_{A(N)}^{02}-W_{A(N)}^{20}\big)
    \label{Rctbarp}\\
R_{L\bar{T}'}^{(N)}&\equiv \mathrm{Im}\big(W_{A(N)}^{32}-W_{A(N)}^{23}\big)
\,.\end{align}

The semi-inclusive responses 
\begin{equation}
    \begin{aligned}
        R_{CT}^{(N)},\,R_{LT}^{(N)},\,R_{C\bar{T}'}^{(N)},\,R_{L\bar{T}'}^{(N)}& \propto \cos\phi_p\\
        R_{TT}^{(N)}&\propto \cos(2\phi_p) 
    \end{aligned}
\end{equation}
are proportional to the cosine of the azimuthal angle \(\phi_p\) of the detected nucleon and vanish when integrated over the full nucleon phase-space.

\section{Delta propagator and adopted form factors}
\label{app:FF}
The \(\Delta\) propagator appearing in the MEC \eqref{eq:MEC}, with momentum \(p\) and mass \(M_\Delta\) is defined as follows:
\begin{equation}
    G_{\alpha \beta}(p)=\frac{\mathcal{P_{\alpha \beta}}(p)}{p^2-M_\Delta^2+iM_\Delta\Gamma_\Delta(p)}
\,,\label{eq:prop}
\end{equation}
where 
\begin{equation}
    \mathcal{P_{\alpha \beta}}(p) =\sum_{spin}u_\alpha(p)\overline{u}_\beta(p)= -(\slashed p+M_\Delta)\Big[g_{\alpha\beta}-\frac{1}{3}\gamma_\alpha \gamma_\beta-\frac{2}{3}\frac{p_\alpha p_\beta}{M^2_\Delta}+\frac{p_\alpha\gamma_\beta-p_\beta\gamma_\alpha}{3M_\Delta}\Big] \,.
\end{equation}
The free $\Delta$ decay width is given by \cite{Dekker:1994yc}:
\begin{equation}
\Gamma_\Delta(p)=\frac{(4 f_{\pi N \Delta})^2}{12\pi m_\pi^2} \frac{|\mathbf{k}|^3}{\sqrt{p^2}} (m_N + E_k) F(k_{\rm rel}^2)\,,
\end{equation}
where $(4 f_{\pi N \Delta})^2/(4\pi)=0.38$, $p^2$ is the $\Delta$ invariant mass, and $\mathbf{k}$ is the three-momentum of the produced pion or nucleon in the $\Delta$-at-rest frame. This three-momentum is expressed as:
\begin{equation}
\mathbf{k}^2=\frac{1}{4p^2}[p^2-(m_N+m_\pi)^2][p^2-(m_N-m_\pi)^2]\,.
\end{equation} 
Additionally, $E_k=\sqrt{m_N^2 + \mathbf{k}^2}$ represents the associated nucleon energy.\\
To better reproduce experimental data, the additional factor~\footnote{In \cite{Dekker:1994yc}, in the denominator of Eq.~(A5), $\Lambda^2$ appears instead of $\Lambda_R^2$.}
\begin{equation}
F(k_{\rm rel}^2)=\left(\frac{\Lambda_R^2}{\Lambda_R^2-k_{\rm rel}^2}\right)
\end{equation}
is considered, where $k_{\rm rel}^2=(E_k - \sqrt{m_\pi^2 + \mathbf{k}^2})^2-4\mathbf{k}^2$ is the relative four-momentum of the $\pi-N$ system and $\Lambda_R^2=0.95\times m_N^2=0.915^2$ GeV$^2$.

To take into account the particle's structure, we applied strong form factors to each interaction vertex involving a pion. In the $\pi NN$ vertex, with a pion carrying a four-momentum \(k\), the $F_{\pi NN}$ hadronic form factor  is inserted
\begin{equation}
    F_{\pi NN}(k^2)=\frac{\Lambda_\pi^2-m_\pi^2}{\Lambda_\pi^2-k^2} \, ,
\end{equation} 
where $\Lambda_\pi=1300$ MeV. The same prescription is applied to the $\gamma \pi NN$ vertex.

In the \(\pi N\Delta\) vertex we use $F_{\pi N \Delta}$ \cite{vanFaassen:1983ma}
\begin{equation}
    F_{\pi N\Delta}(k)=\frac{\Lambda_{\pi N\Delta}^2}{\Lambda_{\pi N \Delta}^2-k^2}\,,
\end{equation}
with $\Lambda_{\pi N \Delta}=1150$ MeV.

In the weak pionic current, the \(F_1^V=F_1^p-F_1^n\) form factor is used to describe the coupling with the weak boson \(W\), accordingly to the weak nucleon current.
The \(\pi \pi NN\) interaction vertex is modeled using the \(F_\rho\) form factor, which reflects the rho dominance in this type of interaction \cite{Hernandez:2007qq}. The PCAC condition requires the axial ``contact'' contribution to be described by the same form factor, reminiscent of the rho-meson propagator carrying the relative pion momenta, here denoted as \(k\)
\begin{equation}
    F_\rho(k^2)\equiv \frac{1}{1-k^2/m_\rho^2} \qquad m_\rho = 775.8 \text{ MeV} \, .
\end{equation}

\section{Isospin algebra for semi-inclusive processes }
\label{app:Isospin}

To illustrate the computation of the isospin part of the semi-inclusive tensor, an example is examined in detail. The detected particle is assumed to be a proton. Thus every final state involving a proton is taken into account. 
For simplicity the \(\nu\)CC weak pion-in-flight direct contribution is analyzed. The corresponding current isospin operator is 
\begin{equation}
    I_{V_+}=- \big(\tau_3^{(1)}\tau_+^{(2)}-\tau_+^{(1)}\tau_3^{(2)} \big)\qquad \tau_+=\tau_1+\tau_2\;.
\end{equation}

Then
\begin{equation}
\sum_{\substack{{t_1,t_2}\\{\text{one proton}}}}\langle t_{1'} t_{2'} |I_{V_+}|{t_1 t_2}\rangle \langle t_1 t_2 |\big(I_{V_+}\big)^\dagger|t_{1'} t_{2'}\rangle=\sum_{\text{one proton}} \langle t_{1'} t_{2'}|\,|I_{V_+}|^2|t_{1'} t_{2'}\rangle\,.
\end{equation}
Using the isospin algebra, four different contributions arise: 
\begin{align}
t_{1'}=t_{2'}=-1/2&\rightarrow  \langle t_{1'} t_{2'}|\,|I_{V_+}|^2|t_{1'} t_{2'}\rangle=0\\
t_{1'}=t_{2'}=+1/2&\rightarrow  \langle t_{1'} t_{2'}|\,|I_{V_+}|^2|t_{1'} t_{2'}\rangle=8\\
t_{1'}\neq t_{2'}&\rightarrow  \langle t_{1'} t_{2'}|\,|I_{V_+}|^2|t_{1'} t_{2'}\rangle=4\,.
\end{align}
Finally, the two non-vanishing configurations contribute to the semi-inclusive cross section with one proton in the final state, yielding
\begin{equation}
     \sum_{\text{one proton}} \langle t_{1'} t_{2'}|\,|I_{V_+}|^2|t_{1'} t_{2'}\rangle=16\,.
\end{equation}
Similar arguments hold for the other contributions, in the direct and exchange terms. Note that the latter are evaluated by changing the order of the state associated with the current with \(p_1 \leftrightarrow p_2\), namely \(|t_{2'} t_{1'}\rangle\).

In the electron scattering case, only the \(\Delta\) diagrams contribute to the \(pp\) and \(nn\) final states, and in semi-inclusive computations just one of the two must be included. In contrast, in the CC neutrino scattering there is always a proton in the final state, so for semi-inclusive processes with a detected final proton all possible configurations contribute.

\section{\(2p2h\) semi-inclusive phase space}
\label{app:PS}

In this Appendix we illustrate how the nine-dimensional integral \eqref{eq:WmunuN} can be reduced to a five-dimensional integral. To avoid the complications related to the explicit expressions of the two-body currents, we focus on the calculation of the phase space integral, which simply corresponds to setting $w^{\mu\nu}_{2p2h}=1$ in Eq. \eqref{eq:WmunuN}, neglecting a factor \({V}/{(2\pi)^9}\).
The integration over the particle \(\bf p_2\) is solved exploiting the spatial momentum conservation Dirac
 $\delta^3$.
Then, the energy conservation \(\delta\) is exploited to perform the integration over one hole polar angle: $\theta_{h_2}$  is chosen. Note that in the following formulas, to improve readability, the bold notation stands for three-vectors, while the normal font indicates the modulus of a vector or a scalar quantity.

Thus the semi-inclusive phase space integral reads
\begin{equation}
\begin{aligned}
    PS_{N}(\omega,q,\bf p_1)=& p_1^2m_N^4\int_{h_i<p_F}  \frac{  \ud{\bf h_1}\ud{\bf h_2}}{E_{h_1}E_{h_2}E_{p_1}E_{p_2}}\delta(E_{p_1}+E_{p_2}-\omega -E_{h_1} -E_{h_2}) \\
    =&\int  \ud h_1\ud h_2 \ud \phi_{h_1} \ud\phi_{h_2} \ud \!\cos{\theta_{h_1}} \sum_{\bar{\theta}_{h_2}}\frac{\sin{\bar\theta_{h_2}}}{R\sin{\gamma}}\frac{ m_N^4 h_1^2  h_2  p_1^2}{E_{h_1} E_{h_2} E_{p_1}}\,.
    \label{eq:Fps-semi}
\end{aligned}
\end{equation}
The values $\bar{\theta}_{h_2}$ are the outcomes of the analytical solution of the energy Dirac $\delta$:
\begin{equation}
\begin{aligned}
    \int& \ud{\cos\theta_{h_2}}\delta(E_{p_1}+E_{p_2}-\omega-E_{h_1}-E_{h_2})=\\
    &=\int \ud{\cos\theta_{p_1}}\frac{E_{p_2}}{h_2}\delta(F-A\cos\theta_{h_2}-B\cos{\theta_{h_2}})=\sum_{\bar{\theta}_{h_2}}\frac{E_{p_2}\sin{\bar\theta_{h_2}}}{h_2 R\sin{\gamma}}\,.
\end{aligned}
\end{equation}
Explicitly, the $\bar\theta_{h_2}$ values are obtained by solving 
\begin{equation}
     \begin{system}
  \sin\theta_{h_2} = \frac{F-A\cos\theta_{h_2}}{B} \qquad \theta_{h_2} \in [0,\pi]\\
      \sin^2\theta_{h_2}+\cos^2\theta_{h_2} = 1 \\
\end{system} \quad \rightarrow \quad \sin\bar{\theta}_{h_2}=\frac{B\cos\gamma\pm sign(B)A\sin\gamma }{R}\,,
\end{equation}
where
\begin{equation}
    A=q+h_1\cos\theta_{h_1}-p_1\cos\theta_{p_1} \qquad B=h_1   \sin\theta_{h_1}\cos(\phi_{h_1}-\phi_{h_2})-p_1\sin\theta_{p_1}\cos(\phi_{h_2}-\phi_{p_1}) 
\end{equation}
\begin{equation}
R=\sqrt{A^2+B^2} \qquad F=\frac{p_2^2-({\bf q+h_1-p_1}^2)-h_2^2}{2h_2}\qquad \cos{\gamma}=\frac{F}{R}\,.  
\end{equation}
Due to the definition range of \(\theta_2\), a polar angle, allowed solutions admit $\sin\bar{\theta}_{h_2}\ge 0$ only. 

\clearpage
\section{Semi-inclusive results}
\label{app:3D}

In this Appendix we present some three-dimensional plots, representing the six-fold differential cross section defined in Eq.~\eqref{eq:xsec-semi1}, which are  discussed in Section~\ref{sec:Results}.

\begin{figure} [hp!]
    \centering    
   \includegraphics[width=1\linewidth]{ 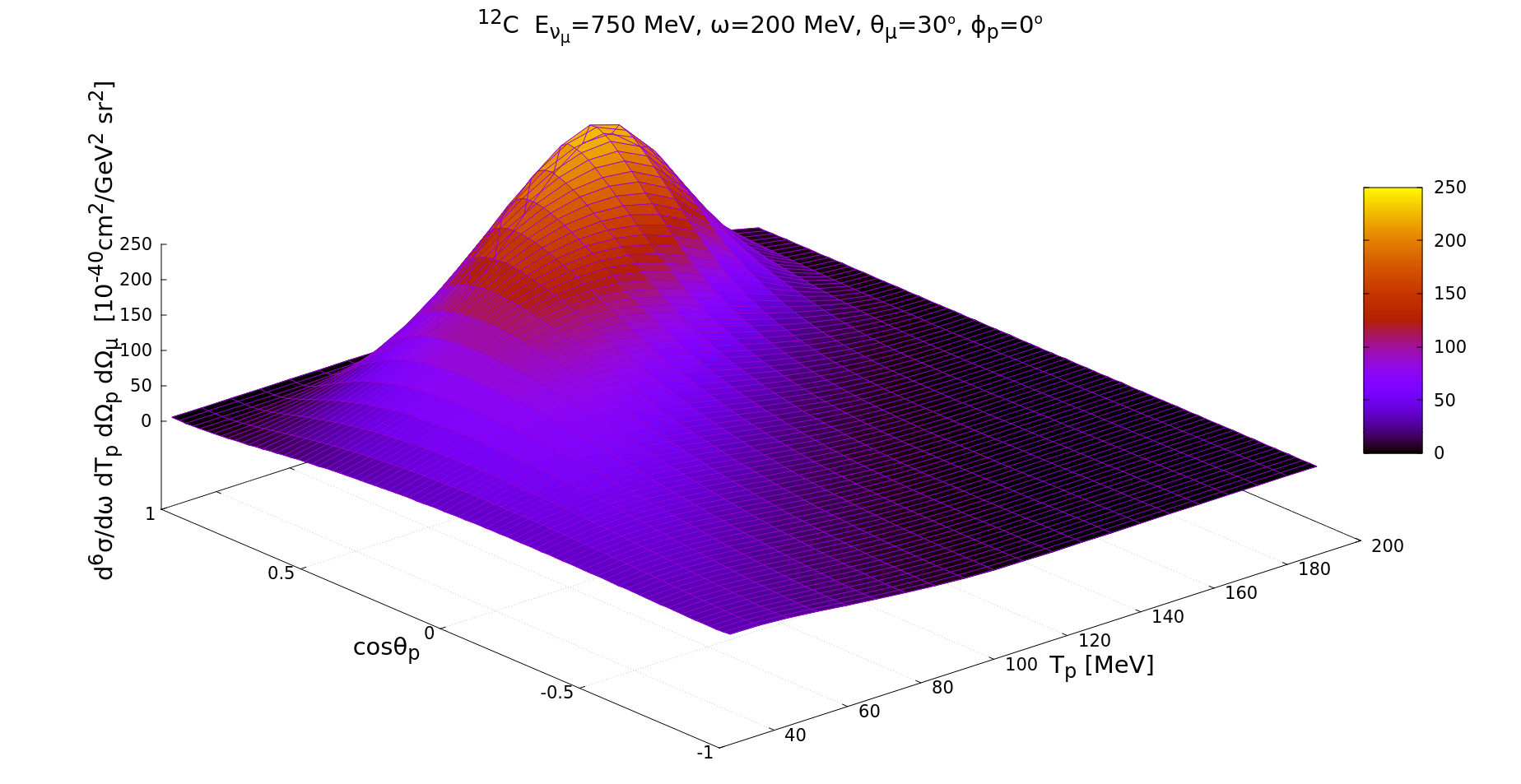}
   \includegraphics[width=1\linewidth]{ 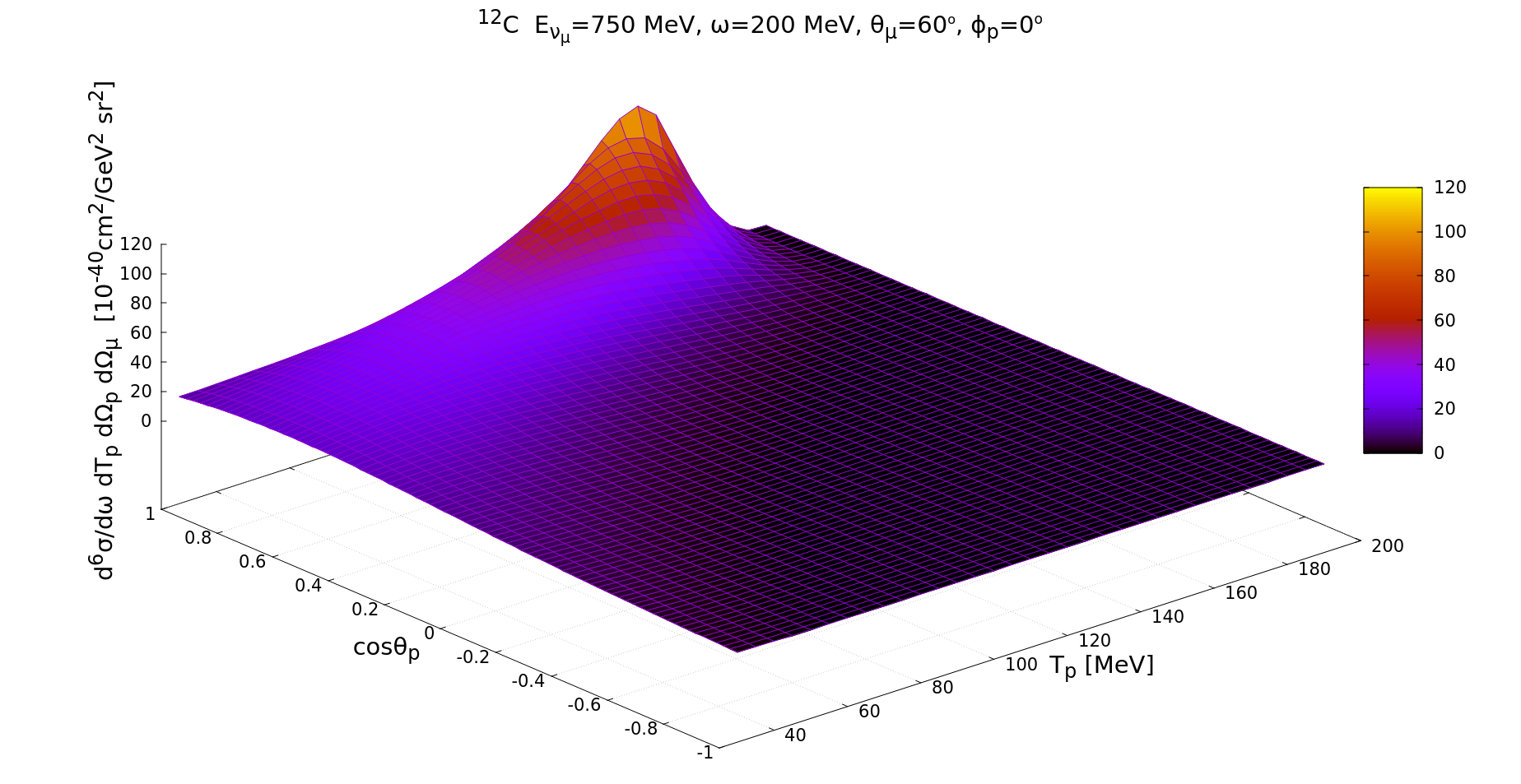}
    \begin{small}\caption{
    Semi-inclusive $\nu_\mu$-$^{12}C$ sixth differential cross section for the \(2p2h\) channel, computed at  \(E_{ \nu_\mu}=750\) MeV and  \(\omega= 200\) MeV. The scattering angle \(\theta_\mu =30\)° (top), \(60\)° (bottom) is fixed, as well as the azimuthal final proton angle \(\phi_p= 0\)°. The cross section is displayed as a function of the polar angle \(\theta_p\) and kinetic energy \(T_p\) of the final proton. The momentum transfer in the two configurations is \({\bf |q|}\simeq391;\,670\) MeV respectively.}
    \label{fig:3D-30-60deg}
    \end{small}
    
\end{figure}
\clearpage
\begin{figure}[hp!]
    \centering
    \includegraphics[width=.75\linewidth]{ 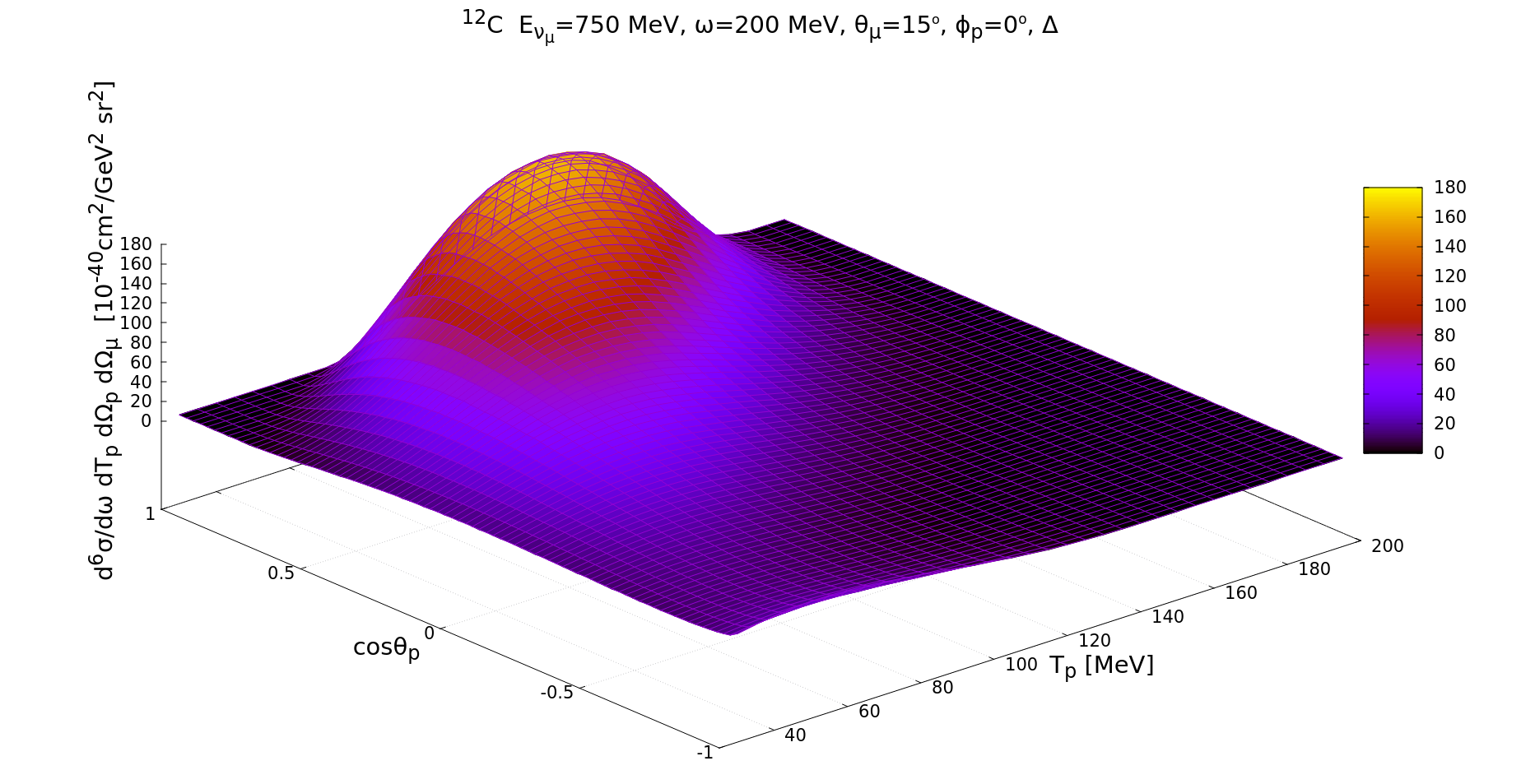}\\
    \includegraphics[width=.75\linewidth]{ 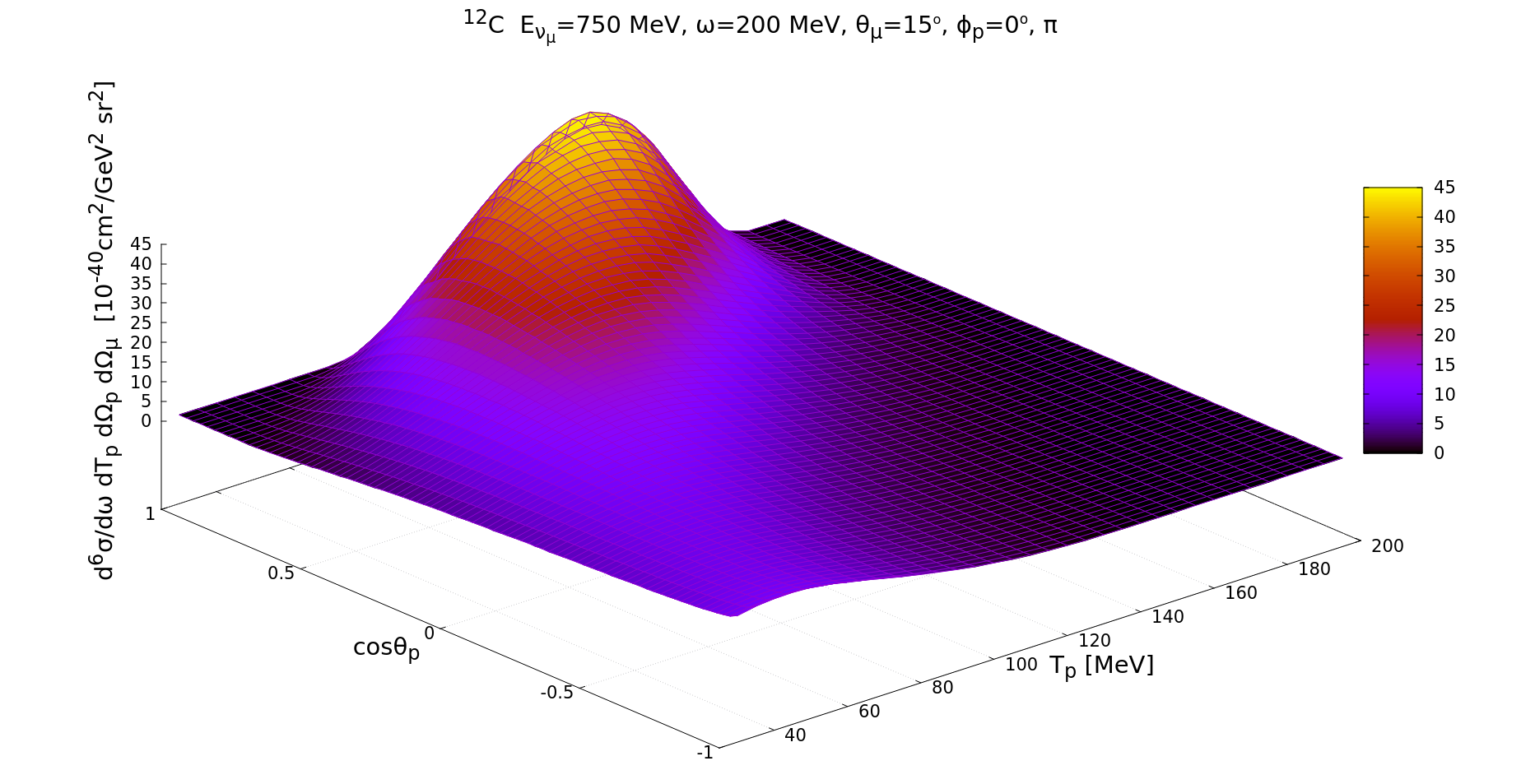}\\
    \includegraphics[width=.75\linewidth]{ 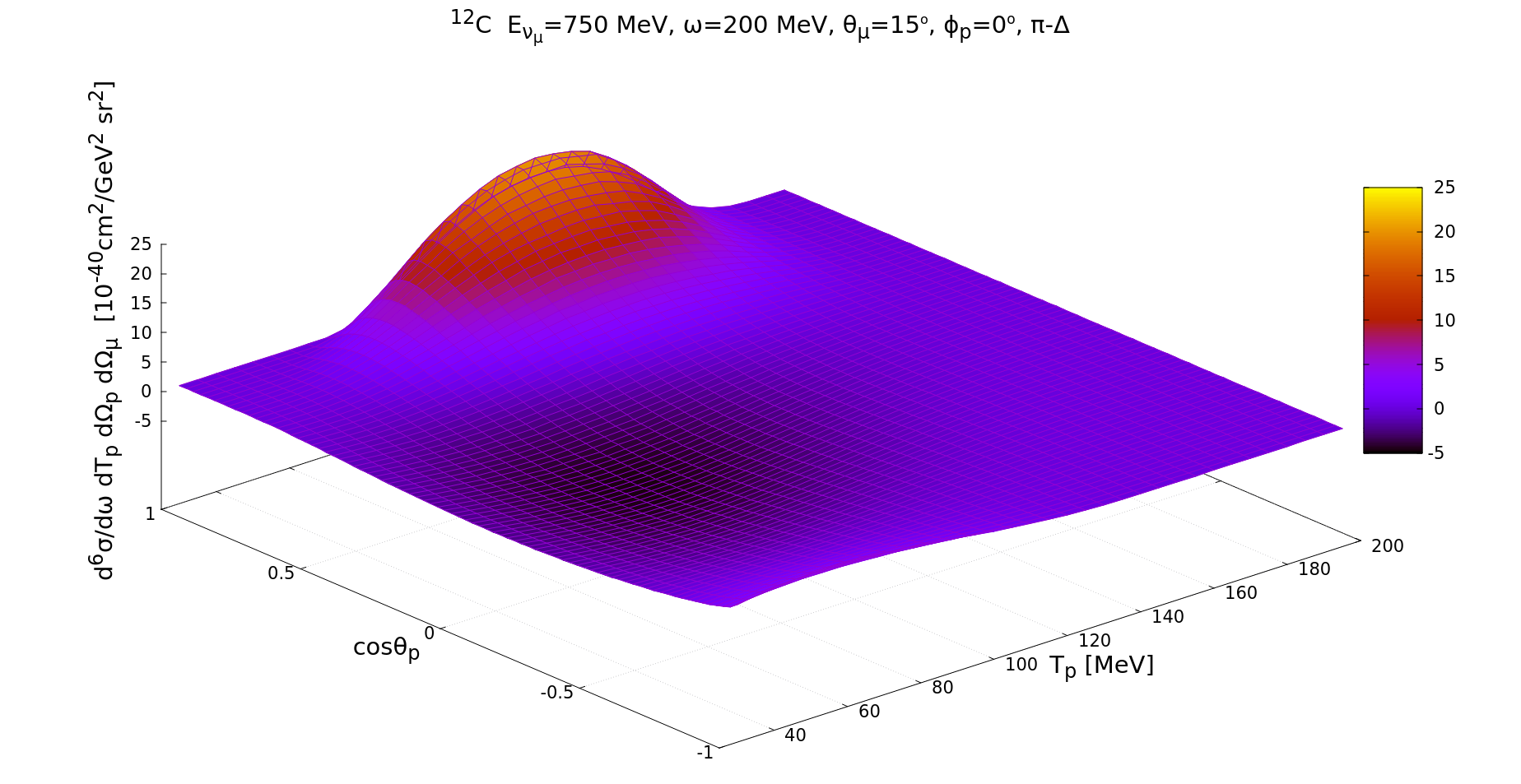}
    \caption{
      Semi-inclusive $\nu_\mu$-$^{12}C$ sixth differential cross section for the \(2p2h\) channel, computed at  \(E_{ \nu_\mu}= 750\) MeV and \(\omega= 200\) MeV. The scattering angle \(\theta_\mu = 15\)° is fixed, as well the azimuthal final proton angle \(\phi_p= 0\)°. The cross section is displayed as a function of the polar angle \(\theta_p\) and kinetic energy \(T_p\) of the final proton. The \(\Delta\) (top), the pionic (center) contributions and the interference between them (bottom) are displayed separately.}
    \label{fig:3D-60deg_study}
\end{figure}
\clearpage
\begin{figure}[hp!]
    \centering
    \includegraphics[width=1\linewidth]{ 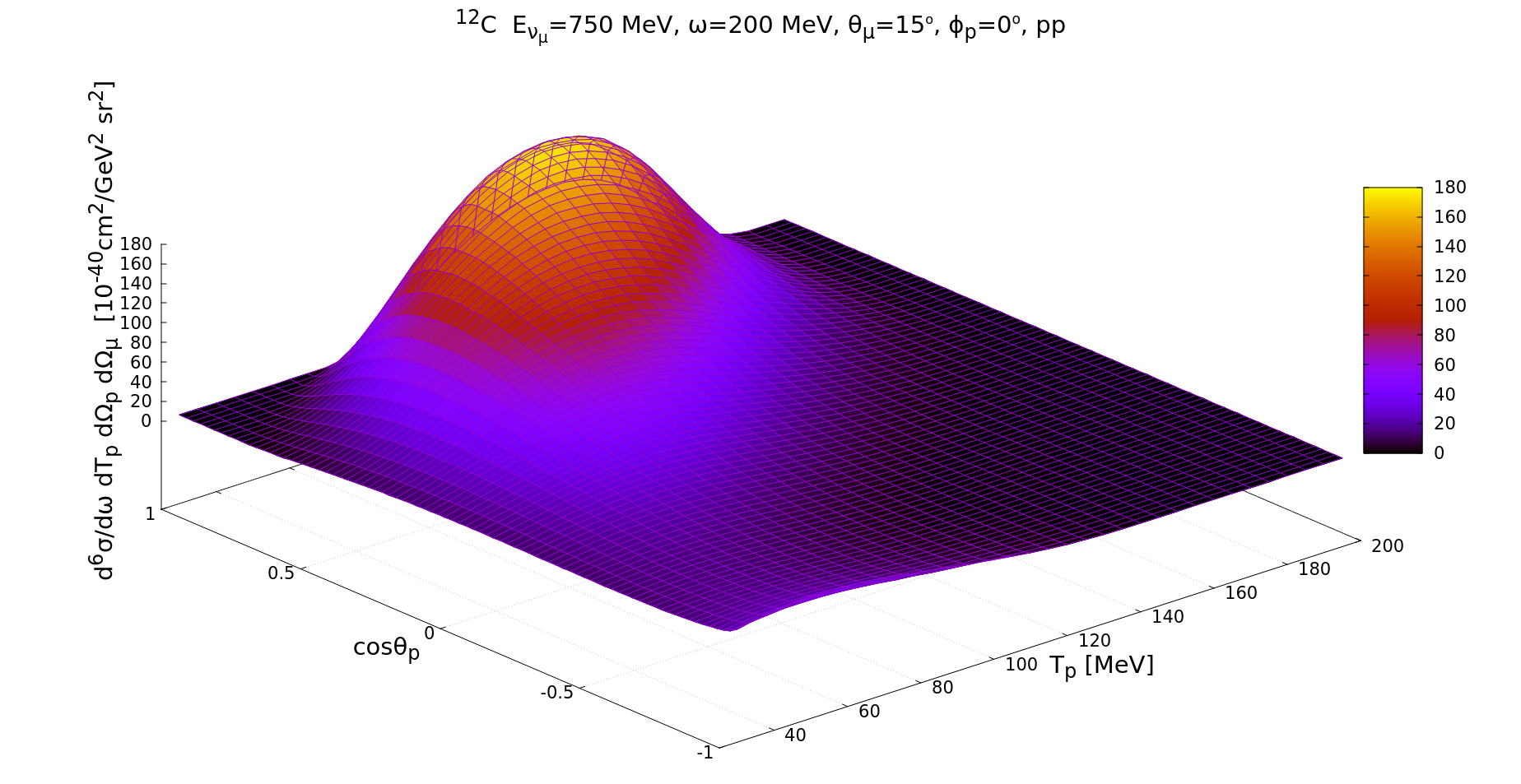}\\ \includegraphics[width=1\linewidth]{ 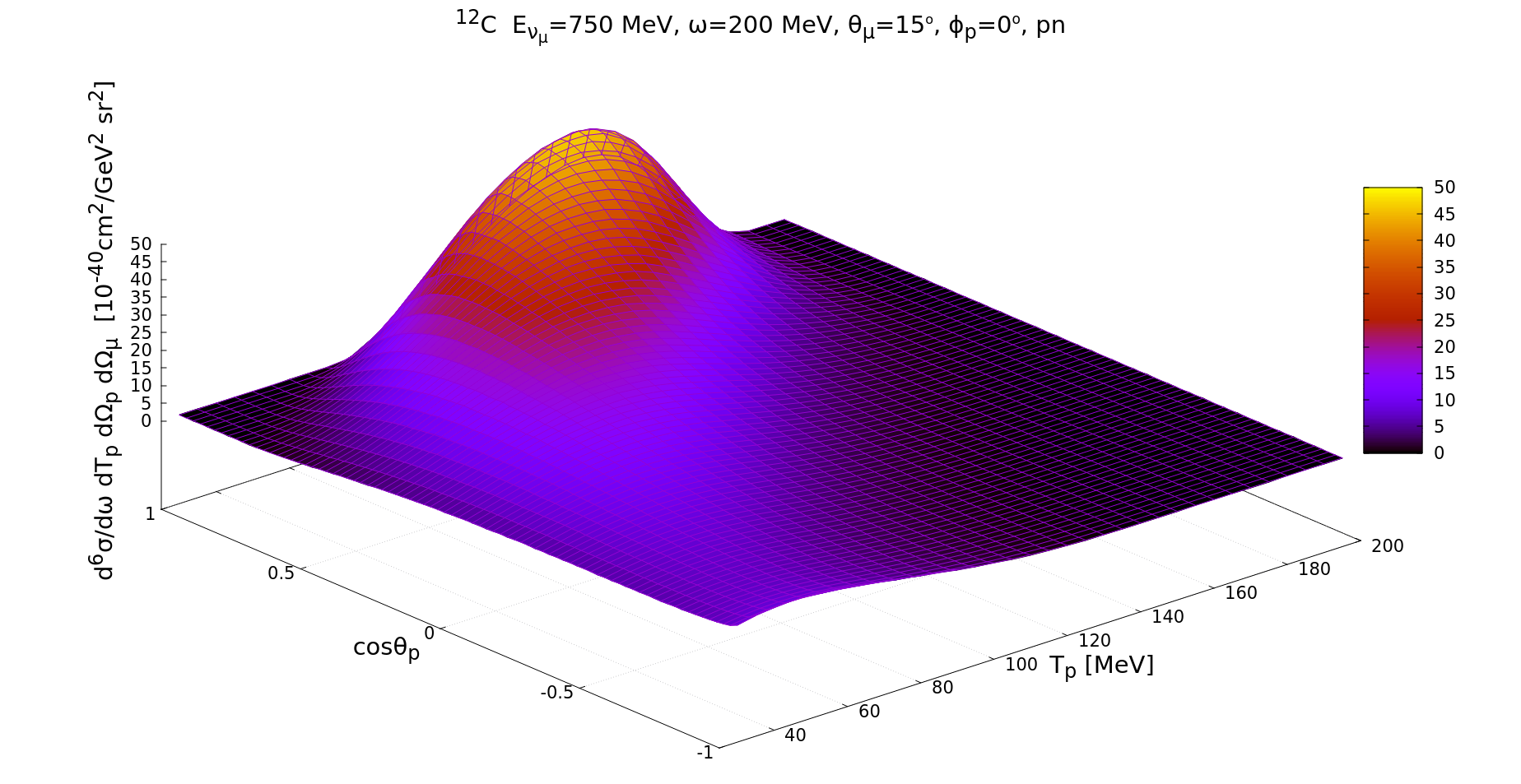}
    \caption{Semi-inclusive $\nu_\mu$-$^{12}C$ sixth differential cross section for the \(2p2h\) channel, computed at  \(E_{ \nu_\mu}= 750\) MeV and  \(\omega= 200\) MeV. The scattering angle \(\theta_\mu =15\)° is fixed, as well the azimuthal final proton angle \(\phi_p= 0\)°. The cross section is displayed as a function of the polar angle \(\theta_p\) and kinetic energy \(T_p\) of the final proton. The two plots correspond to the contributions of the  \(pp\) (top) and \(pn\) (bottom) final states.}
    \label{fig:3D-15pp_pn}
\end{figure}

\section{Laboratory frame}
\label{App3}
All the computation are performed in the q-system, a frame in which the three-momentum transfer \({\bf q}\) plays a key role. In fact the \(z\)-axis is chosen to be parallel to \({\bf q}\): this choice reduces the computational effort noticeably, especially in the leptonic and hadronic tensors' evaluation. 
Furthermore, the azimuthal invariance, derived by the assumed spherical symmetry of the nucleus, is clearly understood with respect to the three-momentum transfer. 

As a matter of fact, due to the neutrino energy broad flux, the momentum transfer is not an observable quantity, as it is not accessible directly. Thus, experimental results are based on initial and final physical quantities, such as particles' momentum and energy. In a neutrino experiment, the incident neutrino energy, and so the momentum, is not known, so that the momentum transfer is also unknown. Nevertheless, the incident neutrino direction is known: in the laboratory frame, the \(z\)-axis is chosen to be parallel to the incident neutrino momentum.

The incident and the final momentum of the leptons interacting in the leptonic vertex define a plane, denoted the scattering plane, which typically corresponds to the \(xz\) plane, in both the two frames, the q-system and the laboratory frame.

The momentum of the ejected nucleon, for simplicity assumed to be the one generated in the hadronic interaction vertex and not through the FSI mechanism, forms a plane with the \(z\)-axis, called the reaction plane.

In our approach, the hadronic system is entirely described in the q-system. Consequently, to compute physical observables in the laboratory frame, a rotation must be applied. In fact, the reaction planes do not coincide in the two frames, and the rotation matrix connecting the two frames involves both polar and azimuthal angles. The magnitude of the momentum \(|{\bf p_N}|\) is clearly not affected by the rotation and is the same in the two frames. In contrast, the angular components of the final proton momentum exhibit a strong dependence on the specific frame. The rotation matrix connecting the two frames is
\begin{equation}
    R(\theta_q)\equiv \left( \begin{matrix}
        \cos \theta_q & 0 & -\sin \theta_q\\ 0 & 1 & 0 \\ \sin \theta_q & 0 & \cos \theta_q
    \end{matrix}\right)\,,\qquad {\bf v}^L=R(\theta_q){\bf v}^q\,,
\end{equation}
where \(\theta_q>0\) is the angle between the incident lepton momentum \({\bf k}\) and the momentum transfer \({\bf q}\),
\({\bf v}^L\) represents a generic three-vector in the laboratory frame, whereas \({\bf v}^q\) is the same vector in the q-system. It is important to note that \({\bf k'}\) always lies in the half-plane with a positive abscissa in both frames. The rotation has a significant effect on the calculation, as it couples the polar and azimuthal angles. This mixing is particularly relevant in the evaluation of the hadronic tensor, when the nuclear responses or the cross-section are presented as functions of hadronic quantities given in the laboratory frame.

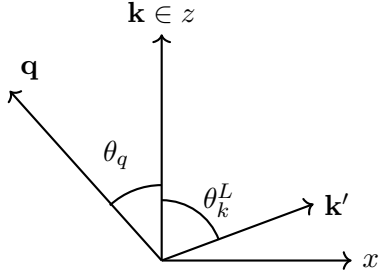
\begin{figure}
    \centering  \begin{tikzpicture}
  \draw[thick,->] (0,0) -- (2,0.75) node[right] {${\bf k'}$};
  \draw[thick,->] (0,0) -- (-2,2.25) node[above right] {${\bf q}$};
  \draw[thick,->] (0,0) -- (0,3) node[above] {${\bf k}  \in z$};
  \draw[thick,->] (0,0) -- (2.5,0) node[right] {$x$};

  \draw [thick,decorate] (0,0.8) arc (90:20:0.8cm);
  \draw [thick,decorate] (0, 1) arc (90:133:1cm); 

  \node at (.75, 0.8) {${\theta_{k}^L}$};
  \node at (-0.6, 1.4) {${\theta_{q}}$};
\end{tikzpicture}

    \caption{Schematic representation of the scattering plane in the laboratory frame. \(\theta_q\) is the angle between the incident lepton momentum \({\bf k}\) and the momentum transfer \({\bf q}\), while \(\theta_k^{L}\) is the scattering angle. Note that \(0^\circ<\theta_q<90^\circ\) for kinematical constraints, stating \({|\bf k|>|\bf k'|}\) and \(\bf q=k-k'\) .}
    \label{fig:rotation}
\end{figure}
\bibliography{biblio}

\end{document}